\documentclass[reprint,amsmath,amssymb,prl,aps]{revtex4-2}
\usepackage[pdftex]{graphicx}
\usepackage{color}
\usepackage{dcolumn}
\usepackage{bm}
\usepackage{physics}
\usepackage{comment}
\usepackage{hyperref}
\hypersetup{
setpagesize=false,
colorlinks=true,
linkcolor=blue,
citecolor=red,
}

\makeatletter
\let\MYcaption\@makecaption
\makeatother
\usepackage{subcaption}
\makeatletter
\let\@makecaption\MYcaption
\makeatother
\begin{document}


\title{Nonreciprocal electron hydrodynamics under magnetic fields: \\applications to nonreciprocal surface magnetoplasmons}

\author{Ryotaro Sano}
\email[]{sano.ryotaro.52v@st.kyoto-u.ac.jp}
\author{Riki Toshio}%
\author{Norio Kawakami}
\affiliation{%
 Department of Physics, Kyoto University, Kyoto 606-8502, Japan
}%

\date{\today}

\begin{abstract}
Recent experiments have elucidated that novel nonequilibrium states inherent in the so-called hydrodynamic regime are realized in ultrapure metals with sufficiently strong momentum-conserving scattering. In this letter, we formulate a theory of electron hydrodynamics with broken inversion symmetry under magnetic fields and find that novel terms emerge in hydrodynamic equations which play a crucial role for the realization of the nonreciprocal responses. Specifically, we clarify that there exist a novel type of nonreciprocal collective modes dubbed \textit{nonreciprocal surface magnetoplasmons} arising from an interplay between magnetic fields and the orbital magnetic moment. We reveal that these nonreciprocal collective modes indeed give rise to the nonreciprocity in magneto-optical responses such as the reflectivity. The physics discussed here will bridge the two important notions of magnetoplasmonics and nonreciprocity in electron hydrodynamic materials with inversion symmetry breaking and magnetic fields.

\end{abstract}

\maketitle


\textit{Introduction---}
The past decades have seen a profound growth of interest in nonreciprocal responses~\cite{Tokura2018NonreciprocalRF}, which are characterized by directional transport and propagation of quantum particles, such as electrons, photons, magnons and phonons. The responses are known to occur in material systems with broken inversion symmetry, as exemplified by the diode in semiconductor p-n junction and the natural optical activity in chiral materials~\cite{Ma_2015}. Especially, a particular type of nonreciprocal responses occurs ubiquitously when the time-reversal symmetry (TRS) is further broken by applying a magnetic field or with spontaneous magnetization: typical examples include nonreciprocal plasma modes and magnetochiral effects in chiral magnets~\cite{Yokouchi2017}. In addition, the former has attracted great attention from both thoeretical and experimental viewpoints due to their extraordinary features and potential applications to a terahertz nonreciprocal isolator~\cite{Fan:18} and a perfect unidirectional absorber~\cite{Yu_2012} in the context of plasmonics.
More recently, the role of such nonreciprocity is highlighted in the context of electron hydrodynamics~\cite{geurs2020rectification}.


Electron hydrodynamics is quickly growing into a mature field of condensed matter physics~\cite{Moll2016,Bandurin2016,Bandurin2018,Crossno1058,Sulpizio2019,Berdyugin2019,kumar2017,Gooth2018,vool2020,Molenkamp1994,Braem2018,Gusev2018-2,Gusev2018,Levin2018,Kumar2019, PhysRevResearch.3.013290,PhysRevB.103.125106,PhysRevB.97.121105,PhysRevB.99.235148,PhysRevX.10.021005,Varnavides2020,cook2021viscometry,robredo2021new, PhysRevResearch.3.023160,funaki2021vorticityinduced,Toshio}.
In fact, many pieces of experimental evidence for hydrodynamic electron flow have been reported through unconventional transport phenomena, in various materials such as a 2D monovalent layered metal PdCoO${}_2$~\cite{Moll2016}, monolayer(ML)/bilayer(BL) graphene~\cite{Bandurin2016,Bandurin2018,Crossno1058,Sulpizio2019,Berdyugin2019,kumar2017}, Weyl semimetals WP${}_2$~\cite{Gooth2018}, WTe${}_2$~\cite{vool2020}, GaAs quantum wells~\cite{Molenkamp1994,Braem2018,Gusev2018-2,Gusev2018,Levin2018} and a noncentrosymmetric metal MoP~\cite{Kumar2019}.

Symmetry of crystals gives a new twist to the concept of electron hydrodynamics. 
Indeed, a number of papers have addressed rich and novel phenomena due to crystal symmetries in the last few years~\cite{PhysRevB.97.121105,PhysRevB.99.235148,PhysRevX.10.021005,Varnavides2020,cook2021viscometry,robredo2021new,Toshio,PhysRevResearch.3.013290,PhysRevB.103.125106, PhysRevResearch.3.023160,funaki2021vorticityinduced}. For example, it was revealed that inversion symmetry breaking or TRS-breaking provokes anomalous hydrodynamic transport phenomena such as a generalized vortical effect~\cite{Toshio,PhysRevB.102.205201,PhysRevB.89.075124} and a vorticity induced anomalous Hall effect~\cite{funaki2021vorticityinduced,PhysRevLett.118.226601}, which do not emerge in usual fluids like water.
However, the role of nonreciprocity in electron hydrodynamics has not been well investigated yet. This naturally motivates us to study electron dynamics under magnetic fields with noncentrosymmetric hydrodynamic materials, which may open up various avenues for research on nonreciprocal electron hydrodynamics.

In this Letter, we derive the hydrodynamic equations for noncentrosymmetric metals under magnetic fields, which are correct up to the second order in electric fields and the first order in magnetic fields. We elucidate the appearance of a novel anomalous term in the momentum flux, which induces unprecedented nonreciprocal hydrodynamic flows, as a consequence of an interplay between magnetic fields and the orbital magnetic moment of electron wavepackets. From a symmetry viewpoint, we find that this term can appear only in the systems without TRS and inversion symmetry. Furthermore, as a demonstration of our nonreciprocal electron hydrodynamic theory, we provide an analysis of \textit{nonreciprocal surface magnetoplasmons} and show that they lead to the directional dichroism in the reflectance peaks in Kretschmann configuration.
An estimation of the order of the nonreciprocity is also given with an effective model for strained graphene with a staggered sublattice potential. 

\textit{Formulation---}
We outline how to derive the magnetohydrodynamic equations for noncentrosymmetric metals where TRS is broken by an external magnetic field (For detail, see
Supplemental Materials). We start from the Boltzmann equation which governs the evolution of the electron distribution function $f^\alpha$ for band $\alpha$,
\begin{equation}
  \partial_tf^\alpha+\dot{\vb*{r}}^\alpha\vdot\grad_{\vb*{r}}f^\alpha+\dot{\vb*{p}}^\alpha\vdot\grad_{\vb*{p}}f^\alpha=\mathcal{C}[f^\alpha],
\end{equation}
where $\mathcal{C}$ is the collision integral due to the scattering processes. The semi-classical equations of motion read~\cite{Xiao2010}
\begin{equation}
  \begin{aligned}
    D^\alpha\dot{\vb*{r}}^\alpha&=\tilde{\vb*{v}}^\alpha+\frac{e}{\hbar}\vb*{E}\times\vb*{\Omega}^\alpha+\frac{e}{\hbar}(\tilde{\vb*{v}}^\alpha\vdot\vb*{\Omega}^\alpha)\vb*{B},\\
    D^\alpha\dot{\vb*{p}}^\alpha&=-e\vb*{E}-e\tilde{\vb*{v}}^\alpha\times\vb*{B}-\frac{e^2}{\hbar}(\vb*{E}\vdot\vb*{B})\vb*{\Omega}^\alpha,
  \end{aligned}
\end{equation}
where the electric field $\vb*{E}$ depends on position and time, while the magnetic field $\vb*{B}$ is assumed to depend only on time for simplicity. $\alpha$ and $\tilde{\vb*{v}}^\alpha\equiv\grad_{\vb*{p}}\tilde{\epsilon}^\alpha$ are a band index and the group velocity of Bloch wavepacket, $\tilde{\epsilon}^\alpha=\epsilon_0^\alpha-\vb*{B}\vdot\vb*{m}^\alpha$ is defined by the sum of the band energy and the Zeeman energy for the orbital magnetic moment $\vb*{m}^\alpha$ of quasiparticles. Note that the orbital magnetic moment can be intuitively interpreted as the self-rotation of the Bloch wavepacket~\cite{Xiao2010}. $\vb*{\Omega}^\alpha$ is the Berry curvature of the Bloch electrons in band $\alpha$, defined as $\vb*{\Omega}^\alpha\equiv\grad_{\vb*{k}}\times\vb*{A}^\alpha$ with $\vb*{A}^\alpha\equiv i\braket{u_{\alpha\vb*{k}}}{\grad_{\vb*{k}}u_{\alpha\vb*{k}}}$. Here, we have defined the lattice-momentum $\vb*{p}\equiv\hbar\vb*{k}$.
Due to the lack of inversion symmetry, both $\vb*{m}^\alpha$ and $\vb*{\Omega}^\alpha$ are allowed to have nonzero values for any $\vb*{p}$.
In addition, $D^\alpha\equiv1+(e/\hbar)\vb*{B}\vdot\vb*{\Omega}^\alpha$ is the modified phase-space density of states.

Following the standard approach~\cite{Landau_Kinetics,hydroofelectron_graphene}, the continuity equations for the particle density and the momentum are obtained in the relaxation-time approximation for momentum-relaxing scattering processes as follows:
\begin{align}
  &\partial_tn_{\vb*{B}}+\div\vb*{J}^n_{\vb*{B}}=0\label{pcl0},\\
  &\partial_tP_{\vb*{B},i}+\partial_j\Pi_{\vb*{B},ij}=F_i+\Gamma_i\label{mcl0},
\end{align}
where $n_{\vb*{B}}$ and $\vb*{P}_{\vb*{B}}$ are the particle density and the momentum of electrons under magnetic fields respectively, $\vb*{F}$ is the driving force due to external fields and $\vb*{\Gamma}$ is the momentum-relaxing force. In Eqs.\eqref{pcl0} and \eqref{mcl0}, $\vb*{J}^n_{\vb*{B}}$ and $\Pi_{\vb*{B},ij}$ are the fluxes of particle density and the momentum.

In the hydrodynamic regime, the system reaches the local equilibrium via normal electron-electron scatterings, which conserve the total momentum of the electron system. For this reason, we can assume that the distribution functions are described with the Lagrange multiplier $\vb*{u}$ as follows: $f_{\mathrm{local}}^\alpha(\vb*{r},\vb*{p},t)=[e^{\beta(\tilde{\epsilon}^\alpha(\vb*{p})-\vb*{u}\vdot\vb*{p}-\mu)}+1]^{-1}$, which is referred to as the local equilibrium distribution function. Note that the inverse temperature $\beta$, as well as the chemical potential $\mu$ and the drift velocity $\vb*{u}$ generally depend on position and time.

Under an assumption that the band energy has a parabolic dispersion $\epsilon^\alpha_0=p^2/2m$ for simplicity, we obtain the momentum flux as
\begin{align}
  \Pi_{\vb*{B},ij}&=mn_0u_iu_j+p\delta_{ij}+\frac{e}{\hbar}\epsilon_{jkl}E_kC_{il}\nonumber\\
   &+\frac{e}{\hbar}B_k\left(u_lM_{lk}\delta_{ij}+u_jM_{ik}\right)+\frac{e}{\hbar}B_ju_lO_{il},
\end{align}
where $n_0$ is the particle density in the absence of magnetic field. The momentum flux contains three novel geometrical tensors: $\hat{C}$, $\hat{M}$ and $\hat{O}$. While $\hat{C}$ was first derived in Ref.~\cite{Toshio} and expresses an electric field induced torque especially for chiral systems~\cite{PhysRevResearch.3.023160}, $\hat{M}$ and $\hat{O}$ are new geometrical tensors peculiar to the noncentrosymmetric fluids under magnetic fields, defined as
\begin{align}
  M_{il}&\equiv\frac{m\hbar}{e}\sum_\alpha\int[dp]\frac{\partial m_l^\alpha}{\partial p_i}f_0(\epsilon_0^\alpha),\\
  O_{il}&\equiv-\sum_\alpha\int[\dd{\vb*{p}}]p_ip_k\Omega^\alpha_k\partial_{p_l}f_0(\epsilon_0^\alpha),
\label{geoten}
\end{align}
where we have introduced the notation $\int[\dd{\vb*{p}}]\equiv\int\dd{\vb*{p}}/(2\pi\hbar)^d$. These geometrical terms play a crucial role for the realization of nonreciprocal hydrodynamic flow, which especially includes the nonreciprocal surface magnetoplasmons, as seen below.
From a symmetry viewpoint, we readily find that these terms, which have the form $\hat{\Pi}^B\propto \vb*{B}\vb*{u}$, can appear only in the systems without TRS and inversion symmetry.

Combining the particle conservation law Eq.\eqref{pcl0} and the momentum conservation law Eq.\eqref{mcl0}, we end up with the generalized Euler equation
\begin{align}
  &mn_0\pdv{u_i}{t}+mn_0(\vb*{u}\vdot\grad)u_i+\partial_ip+en_0(E_i+\epsilon_{ikl}u_kB_l)\nonumber\\
  &+\frac{e}{\hbar}M_{ij}\pdv{B_j}{t}+\frac{e}{\hbar}B_k\left[M_{lk}\partial_iu_l+M_{ik}\partial_ju_j+O_{il}\partial_ku_l\right]\nonumber\\
  &=-\frac{mn_0u_i}{\tau},\label{ns1}
\end{align}
where the terms with spatial derivatives of $T$ and $\mu$ are omitted for simplicity. This result indicates that the symmetry lowering of the crystal leads to the emergence of anomalous driving forces proportional to $\hat{M}$ or $\hat{O}$ in the conventional Euler equation~\cite{Landau_Fluid}. For example, one of these forces is proportional to $\partial\vb*{B}/\partial t$, and thus represents the inverse Edelstein effect~\cite{inverse_edelstein}, which is an anomalous electric flow induced by the oscillating magnetic field. On the other hand, the other forces that are proportional to $\partial_iu_j$ are also understood as nonreciprocal forces, which never change their direction under the spatial inversion.
Here, we note that the geometrical tensor $\hat{O}$ becomes zero in 2D-layered systems [see Eq.\eqref{geoten}] because it contains the inner product of momentum and Berry curvature which are generally perpendicular in 2D-systems. In the following analysis, we focus on 2D electron hydrodynamic materials such as ML-graphene for simplicity.

To relate the hydrodynamic theory with an observable current in optical experiments, we further need to introduce the so-called local current~\cite{Xiao2006} when we consider the magneto-optical responses,
\begin{equation}
  \vb*{j}=-e\int[\dd{\vb*{p}}]D^\alpha\dot{\vb*{r}}^\alpha f^\alpha+\curl\int[\dd{\vb*{p}}]\vb*{m}^\alpha f^\alpha.
\end{equation}
After straightforward calculations, we obtain the local current in terms of hydrodynamic variables in the same conditions as in Eq.\eqref{ns1},
\begin{equation}
  \vb*{j}=-en_0\vb*{u}-\frac{me^2}{\hbar}\vb*{E}\times({}^t\hat{D}\vb*{u}+{}^t\hat{Y}\vb*{B})+\frac{e}{\hbar}\grad\times ({}^t\hat{M}\vb*{u}),\label{loc1}
\end{equation}
where the first term is a familiar part appearing in the conventional hydrodynamics, and the others are anomalous parts reflecting the symmetry lowering of the fluids. The terms proportional to $\hat{D}$ and $\hat{Y}$ can be understood as hydrodynamic counterparts of two types of the Hall current known in the ohmic or the ballistic regime~\cite{Sodemann2015,Morimoto}. 

\textit{Nonreciprocal surface plasmons in 2D electron hydrodynamic materials---}
Collective behavior is one of the most intriguing aspects of the hydrodynamic approach. In fact, a number of previous works have applied hydrodynamics to analyze collective modes~\cite{Gorbar2018,helicons,PhysRevB.103.115402,PhysRevLett.118.127601,chiralberryplasmon2,Song4658,surface_plasmon1}, particularly plasma modes~\cite{PhysRevB.103.115402,PhysRevLett.118.127601,surface_plasmon1,Song4658,chiralberryplasmon2}. We emphasize that our theory highlights both the effects of crystal symmetry and the geometrical effect of Bloch wave function, which have been rarely addressed in these previous works. In the study of collective modes, we assume that deviations of the local thermodynamic parameters from their equilibrium values are small, justifying the use of linearized hydrodynamic equations. The components of linear conductivity tensor are obtained by looking for a solution in the form of plane waves, i.e. $\vb*{u}=\tilde{\vb*{u}}e^{i\vb*{q}\vdot\vb*{r}-i\omega t}$ together with similar expressions for other oscillating variables.

In the following analysis, we consider surface plasmons in 2D electron hydrodynamic materials such as monolayer graphene~\cite{Bandurin2016,Bandurin2018,Crossno1058,Sulpizio2019,Berdyugin2019,kumar2017}. Surface plasmons originate from the collective excitations of electrons coupled to the electromagnetic field at an interface between a dielectric material and a metal.
The equation which determines the dispersion relation of surface plasmons is given by~\cite{diadic_greensfunc,surface_plasmon1,dynamics_of_qp}
\begin{equation}
  \left[\frac{\epsilon_1}{\kappa_1}+\frac{\epsilon_2}{\kappa_2}+\frac{i\sigma_{yy}}{\epsilon_0\omega}\right]\left[\kappa_1+\kappa_2-\frac{i\omega\sigma_{xx}}{\epsilon_0c^2}\right]=\frac{\sigma_{xy}\sigma_{yx}}{(\epsilon_0c)^2}.
\end{equation}
Here $\kappa_i=\sqrt{q^2-\epsilon_i(\omega/c)^2}\ (i=1,2)$, $\epsilon_i$ are the relative dielectric constants in the medium above ($i=1$) and below ($i=2$) the graphene, $\epsilon_0$ and $c$ are the dielectric constant and the speed of light in free space respectively, and $\sigma_{ij}$ are the components of the conductivity tensor. 
From this equation, we readily notice that the conductivity tensor $\hat{\sigma}(\omega,\vb*{q})$ must have the odd part with respect to $\vb*{q}$ in order to realize nonreciprocal magnetoplasmon.

\begin{figure}[t]
  \centering
  \begin{minipage}{0.45\hsize}
        \centering
        \includegraphics[keepaspectratio,scale=0.35]{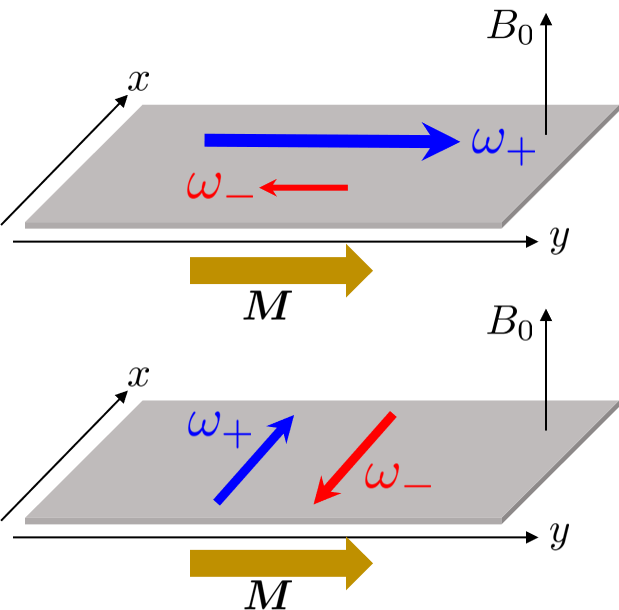}
        \subcaption{}
        \label{fig:spp_total}
      \end{minipage}
      \begin{minipage}{0.45\hsize}
          \centering
          \includegraphics[keepaspectratio,scale=0.35]{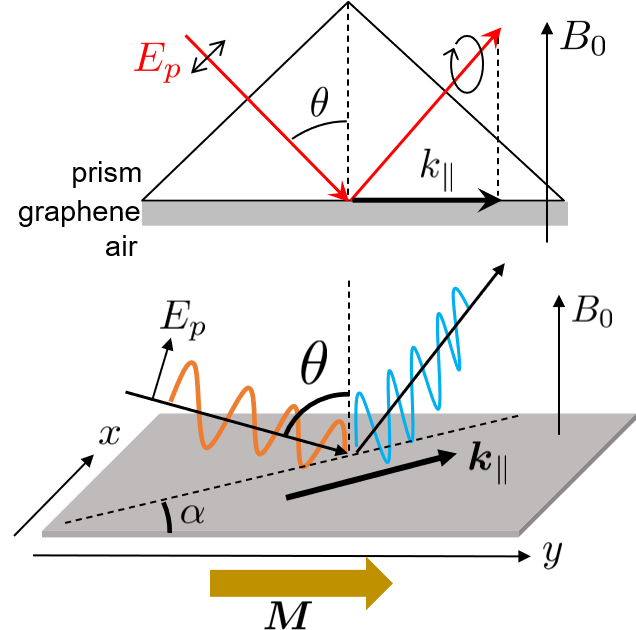}
          \subcaption{}
          \label{fig:Kretschmann}
        \end{minipage}
  \caption{(a) Schematic picture of nonreciprocal surface magnetoplasmons: the blue line is forward propagating, while the red line is backward in the presence of an out-of-plane magnetic field $B_0$. (b) Schematics of Kretschmann configuration: $\mathrm{SiO}_2$ prism ($n_1 = 1.5$) puts on the graphene. 
The direction of the propagation of TM mode incident light is determined by two parameters $\alpha$ and $\theta$.}
\end{figure}

According to symmetry consideration~\cite{Toshio}, the geometrical tensor $\hat{M}$ vanishes in a pseudo-vector constrained in the 2D plane: $M_{ij}=M_i\delta_{jz}$ (see Fig.~\ref{fig:spp_total}). Here we take the vector $\vb*{M}$ along $y$-axis without loss of generality.
In this condition, substituting the solution of Eq.\eqref{ns1} to Eq.\eqref{loc1}, we obtain the components of the linear optical conductivity tensor as follows:
\begin{align}
  \sigma_{xx}&=\sigma_{\mathrm{D}}(1-i\omega\tau)(1+\zeta^2q_y^2)/Z,\\
  \sigma_{xy}&=-\sigma_{\mathrm{D}}[(1+\zeta^2q_x^2)\omega_c\tau+i\zeta q_y(1-i\zeta q_x)(1-i\omega\tau)]/Z\nonumber\\
  &\qquad-me^2Y_{zz}B_0/\hbar,\\
  \sigma_{yx}&=\sigma_{\mathrm{D}}(1+\zeta^2q_x^2)[\omega_c\tau+i\zeta q_y(1+i\zeta q_x)(1-i\omega\tau)]/Z\nonumber\\
  &\qquad+me^2Y_{zz}B_0/\hbar,\\
  \sigma_{yy}&=\sigma_{\mathrm{D}}(1-i\omega\tau)(1+\zeta^2q_x^2)/Z.
\end{align}
Here, $\sigma_{\mathrm{D}}$ is the Drude weight and $\vb*{q}=(q_x,q_y)$ is the in-plane wave vector of the incident light. The denominator is given by
\begin{equation}
  Z\equiv(1-i\omega\tau)(1-i\omega\tau+2i\omega_{\mathrm{c}}\tau\zeta q_y)+(\omega_c\tau)^2(1+\zeta^2q_x^2)
\end{equation}
with $\zeta=M_y/(\hbar n_0)$ and the cyclotron frequency $\omega_{\mathrm{c}}$ of electrons. We find that the nonreciprocity appears most strongly when $\vb*{M}$ is parallel to the propagation direction $\vb*{q}$ of surface magnetoplasmons, while the nonreciprocity completely vanishes when $\vb*{M}$ is perpendicular to $\vb*{q}$ (Fig.~\ref{fig:spp_total}). Thus we conclude that not only the existence of applied static magnetic field $B_0$ but also the angle $\alpha$ defined by $\cos\alpha\equiv\vb*{M}\vdot\vb*{q}/\abs{\vb*{M}}\abs{\vb*{q}}$ are the keys for the appearance of nonreciprocity in the dispersion relation of surface magnetoplasmons.

\begin{figure}[htbp]
  \centering
    \includegraphics[keepaspectratio,scale=0.45]{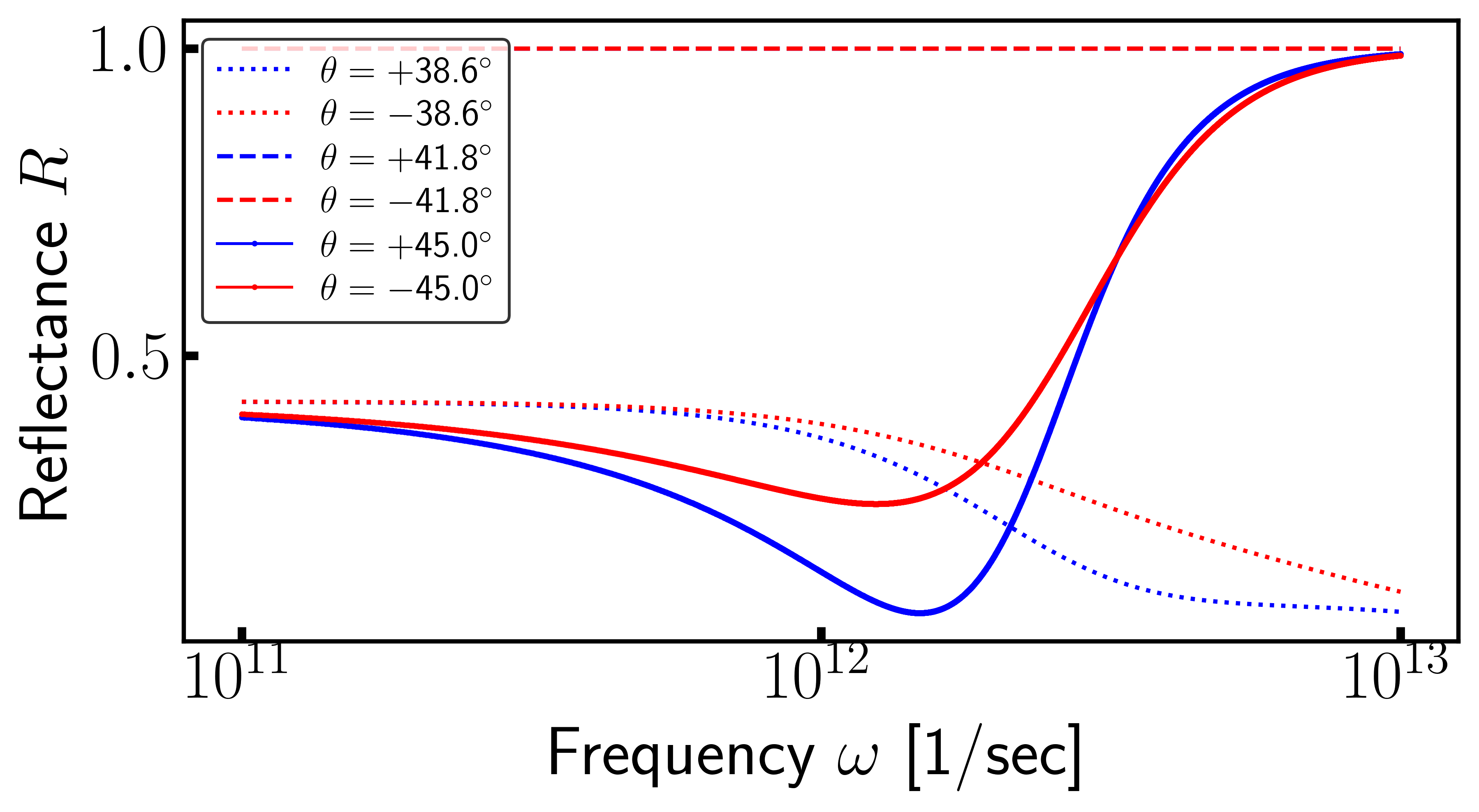}
    \caption{The forward (blue) and backward (red) reflectance as functions of the frequency for three incident angles. We use the parameters; $B_0=10\,\mathrm{T}$, $N_0=8\times10^{13}\,\mathrm{cm^{-2}}$, $\tau=1\times10^{-12}\,\mathrm{s}$, $m=1.3m_{\mathrm{e}}$, $M_{yz}=-8\times10^{-21}\,\mathrm{kg\cdot m/s}$, $Y_{zz}=2.5\times10^{24}\,\mathrm{C\cdot s/kg^2}$, where $m_{\mathrm{e}}$ is the bare electron mass. Here, a siginificant peak splitting appears when $\theta=45^{\circ}$.}
    \label{fig:reflectance}
\end{figure}

Previous works of collective modes in 2D electron hydrodynamic systems have addressed the nonreciprocity in dispersion relation only for the edge modes~\cite{chiralberryplasmon2,Song4658,surface_plasmon1}, but the surface modes have been rarely discussed. This is because these works do not take account of the anisotropy of the systems, which is built in our theory, and thus their theory cannot treat the nonreciprocal surface modes in layered systems due to the strong constrains from the artificial isotropy. Our theory makes clear the role of crystal symmetry and geometrical properties in describing the nonreciprocal surface modes, dubbed \textit{nonreciprocal surface magnetoplasmons}, and characterizes it through magneto-optical signatures such as split peaks in magneto-optical reflectance as shown below (see Fig.~\ref{fig:reflectance}).

\textit{Nonreciprocity in reflectance---}
Next, in order to reveal the role of nonreciprocal surface magnetoplasmons, we consider the magneto-optical responses 
 in Kretschmann configuration, whose schematic picture is dipicted in Fig.~\ref{fig:Kretschmann}. Importantly, we clarify that these modes cause the nonreciprocity in the magneto-optical reflectance. 
  Notice that the polarization plane of the reflective light rotates through the magneto-optical Kerr effect.

After straightforward calculations, we can obtain the reflectance $R$ (Fig.~\ref{fig:reflectance}).
The reflectance has a minimum where the phase matching condition between the incident light and a surface magnetoplasmon is satisfied: namely the horizontal component of momentum of light matches the real part of momentum of surface magnetoplasmon polaritons. This condition can be written
as
\begin{equation}
  k_{\parallel}=k_0n_1\sin\theta_{\mathrm{r}}=\Re{k_{\mathrm{smp}}(\omega)}
\end{equation}
where $k_0$ $(k_{\mathrm{smp}})$ is the wave vector of the initial light (surface magnetoplasmon polariton), $n_1$ is the refractive index of the prism and $\theta_{\mathrm{r}}$ is the resonant angle.

\begin{figure}[t]
  \centering
  \includegraphics[keepaspectratio,scale=0.45]{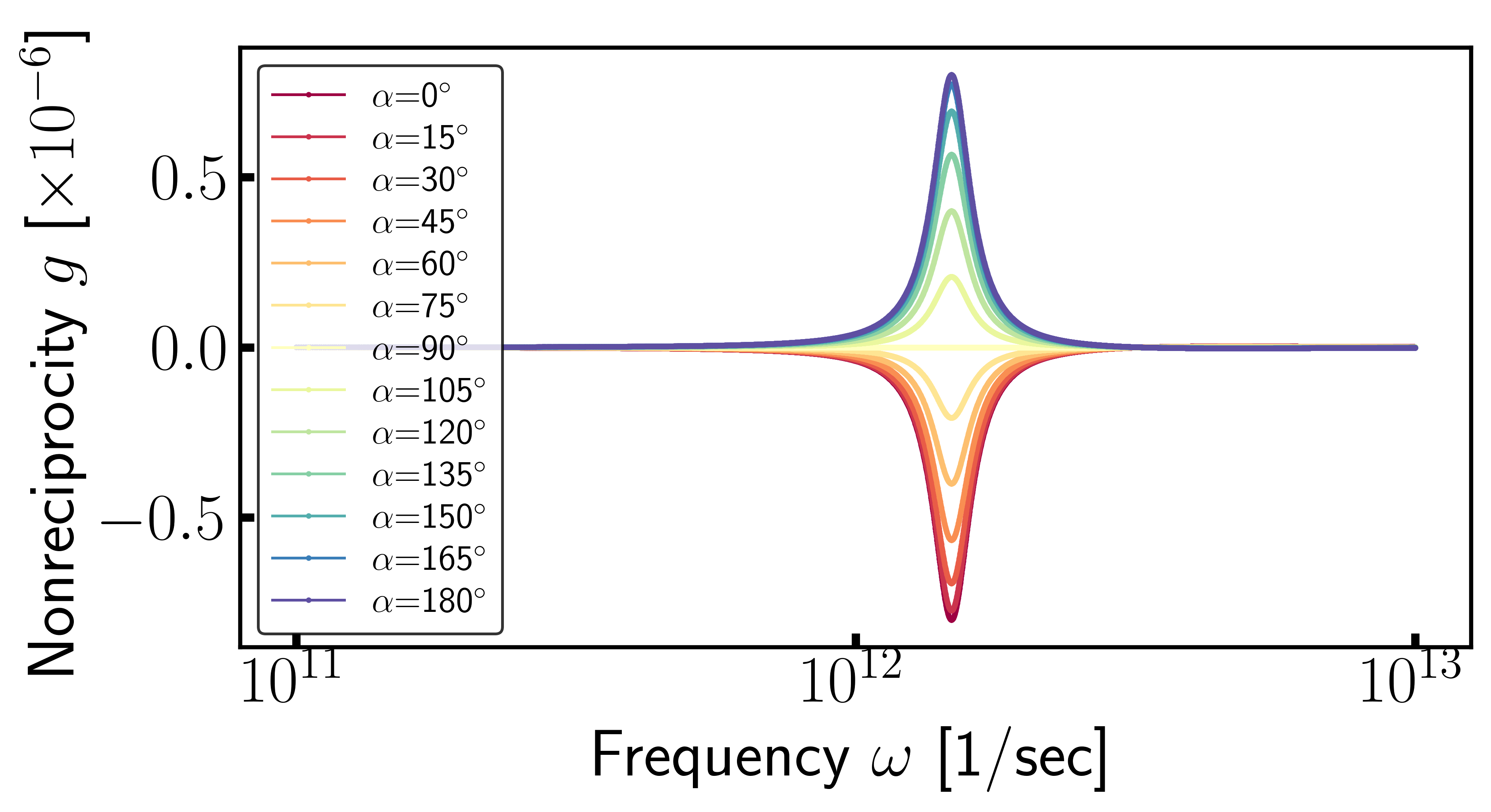}
\caption{Nonreciprocity in the reflectance with altering $\alpha$ from $0^\circ$ to $180^\circ$ with fixed $\theta=45^{\circ}>\theta_{\mathrm{c}}$.}
\label{fig:alpha}
\end{figure}

We are now ready to discuss the nonreciprocity in the reflectance, given by
\begin{equation}
  g=\frac{R(\vb*{k}_\parallel)-R(-\vb*{k}_\parallel)}{R(\vb*{k}_\parallel)+R(-\vb*{k}_\parallel)},
\end{equation}
where $R(\vb*{k}_\parallel)$ and $R(-\vb*{k}_\parallel)$ mean the forward and backward reflectance respectively. We stress that $g$ is most strongly enhanced when $M_y$ is parallel ($\alpha=0$) or antiparallel ($\alpha=\pi$) to the momentum $\vb*{k}_\parallel$ transferrd from electromagnetic radiation, while the nonreciprocity vanishes when $M_y$ is perpendicular to the momentum $\vb*{k}_\parallel$ ($\alpha=\pi/2$) (Fig.~\ref{fig:alpha}). These characteristic behaviors have already been mentioned in the above considerations of nonreciprocal surface mangetoplasmons. 
This indicates that the difference between the forward and backward reflectance stems from the nonreciprocity of the excited magnetoplasmonic modes.

\textit{Model for quantitative estimation---}
Finally, let us consider a simple model for strained graphene with a staggered sublattice potential to estimate the observables quantitatively. As is well-known, the Hamiltonian around the $K$ and $K'$ valley points is given by~\cite{Sodemann2015}
\begin{equation}
  H= vp_x\sigma_y-\alpha vp_y\sigma_x+\alpha s p_y\hat{\bm{1}}_2+\Delta\sigma_z
\end{equation}
where $\alpha=\pm1$ denotes the valley index at $\vb*{P}_{\pm}=\frac{2\pi\hbar}{a}(\pm2/3,0)$ and $\Delta$ is the mass gap. $s$ denotes the strain parameter which is required for a nonzero $M_y$ (see below). This Hamiltonian has a mirror symmetry $\mathcal{M}_y\ (y\rightarrow-y)$ and thus the geometrical tensor $\hat{M}$ must be in the $y$-direction (detailed symmetry consideration in~\cite{Toshio}). By diagonalizing the Hamiltonian, we can obtain the expression of the energy dispersion $\epsilon_0^\alpha=\alpha s p_y+\mathrm{sgn}(\mu)\sqrt{\Delta^2+v^2\vb*{p}^2}$, the Berry curvature and the orbital magnetic moment respectively,
\begin{align}
  \Omega^\alpha&=-\frac{\mathrm{sgn}(\mu)}{2}\frac{\alpha v^2\hbar^2\Delta}{(\Delta^2+v^2\vb*{p}^2)^{3/2}},\\
  m^\alpha&=-\frac{1}{2}\frac{\alpha e\hbar v^2\Delta}{\Delta^2+v^2\vb*{p}^2}.
\end{align}
Here $\mu>0$ $(\mu<0)$ for the conduction (valence) band. Especially in the low carrier and week strain limit, we can approximate the dispersion with an isotropic parabolic form : $\epsilon\simeq\mathrm{sgn}(\mu)\left[\Delta+(\vb*{p}+\vb*{p}^\alpha_0)^2/2m\right]+O(s/v)^2$ where $m=\Delta/v^2$ and $\vb*{p}^\alpha_0=(0,\mathrm{sgn}(\mu)\alpha s\Delta/v^2)$. The hydrodynamic coefficients $\hat{M}$ and $\hat{Y}$ can also be easily calculated in the above limit, and finally we arrive at the following analytic formula : $M_{xz}=0$ and
\begin{align}
  M_{yz}&=-\frac{\mathrm{sgn}(\mu)}{2\pi}\frac{s mv^2p_F^2}{\Delta^2+v^2p_F^2}+O(s/v)^2,\\
  Y_{zz}&=\frac{\mathrm{sgn}(\mu)}{4\pi}\frac{e\hbar v^4\Delta^2}{(\Delta^2+v^2p_F^2)^{5/2}}+O(s/v)^2,
\end{align}
where $p_F=\sqrt{2m(\mu-\Delta)}$ is the radius of the Fermi surface. We can estimate a typical scale of $M_{yz}$ and $Y_{zz}$ for ML-graphene as $-1.4\times10^{-27}\,\mathrm{kg\cdot m/s}$ and $2.5\times10^{24}\,\mathrm{C\cdot s/kg^2}$, using the parameters $v\sim 4.5\times10^5\,\mathrm{m/s}$, $\Delta\sim 1.5\,\mathrm{eV}$, $s/v=0.1$, and $(\mu-\Delta)/\Delta=0.1$ which corresponds to $n_0\simeq 8\times10^{13}\,\mathrm{cm^{-2}}$. Unfortunately, a quantitative estimation shows the nonreciprocity in the reflectance by use of these parameters is of the order of $10^{-6}$ when substituting $B_0=1\,\mathrm{T}$ and is too small to observe in experiments (Fig.~\ref{fig:alpha}). 
However, such a nonreciprocity is expected to be much enhanced in Weyl semimetals or anomalous Hall systems. In the former, the orbital magnetic moment behaves divergently at a Weyl point and thus it is expected that $M_{y}$ has much larger values. On the other hand, in the latter, the spontaneous magnetic order provides a much stronger effective magnetic field which also leads to an enhancement of the nonreciprocity. 
Ideal candidates for the latter could be realized in doped topological insulators such as Cr-doped bismuth telluride~\cite{Chang167} or atomic layered materials (graphene or TMDs) on a magnetic substrate~\cite{PhysRevLett.114.016603,doi:10.1063/1.5001318,Averyanov2018}, both of which satisfy the requirements in our theory. These issues will be left for future investigations.

\textit{Conclusion---}
In summary, we have developed a basic framework of nonreciprocal electron hydrodynamics in noncentrosymmetric metals under magnetic fields, which is composed of a generalized Euler equation Eq.\eqref{ns1} and the hydrodynamic expression of local current Eq.\eqref{loc1}. In addition, we have identified the interplay between the orbital magnetic moment of electron wavepackets and magnetic fields as a geometric origin of nonreciprocity in dispersion relation of surface magnetoplasmons. Furthermore, we have shown that the peak splitting in the magneto-optical reflectance stems from the nonreciprocity in surface magnetoplasmons. 
Our results offer valuable insights into the nonreciprocal response and will pave the way to terahertz plasmonics applications. 

\textit{Acknowledgments---}
We are thankful to Taisei Kitamura, Koki Shinada, Kosuke Nogaki, Akira Kofuji and Hikaru Watanabe for valuable discussions. We also thank Koichiro Tanaka for providing helpful comments from an experimental point of view. This work is partly supported by JSPS KAKENHI (Grants JP20J22612, JP18H01140 and JP19H01838).

\bibliographystyle{apsrev4-2}
\bibliography{ref}
\end{document}



\title{Supplemental Materials for \\ "Nonreciprocal electron hydrodynamics under magnetic fields:\\ applications to nonreciprocal surface magnetoplasmons"}

\author{Ryotaro Sano}
 \email{sano.ryotaro.52v@st.kyoto-u.ac.jp}
\author{Riki Toshio}%
\author{Norio Kawakami}
\affiliation{%
 Department of Physics, Kyoto University, Kyoto 606-8502, Japan
}%

\date{\today}

\maketitle
\renewcommand{\theequation}{S\arabic{equation}}
\onecolumngrid
\section{Formulation}
In this section, we outline how to derive the magnetohydrodynamic equations for noncentrosymmetric metals where TRS is broken by an external magnetic field. We start from the Boltzmann equation,
\begin{equation}
  \partial_tf^\alpha+\dot{\vb*{r}}^\alpha\vdot\grad_{\vb*{r}}f^\alpha+\dot{\vb*{p}}^\alpha\vdot\grad_{\vb*{p}}f^\alpha=\mathcal{C}[f^\alpha],\label{bolzmann}
\end{equation}
where $\mathcal{C}$ is the collision integral due to the scattering processes, given by the sum of the momentum-conserving collision integral $\mathcal{C}^{\mathrm{mc}}$ and the momentum-relaxing collision integral $\mathcal{C}^{\mathrm{mr}}$: $C[f^\alpha]=C^{\mathrm{mc}}[f^\alpha]+C^{\mathrm{mr}}[f^\alpha]$.
The semi-classical equations of motion read~\cite{Xiao2010}
\begin{equation}
\begin{aligned}
  &\dot{\vb*{r}}^\alpha=\tilde{\vb*{v}}^\alpha-\dot{\vb*{k}}^\alpha\times\vb*{\Omega}^\alpha,\\
  &\dot{\vb*{p}}^\alpha=-e\vb*{E}-e\dot{\vb*{r}}^\alpha\times\vb*{B},
\end{aligned}\label{SEOM1}
\end{equation}
where the electric field $\vb*{E}(\vb*{r},t)$ depends on position and time, while the magnetic field $\vb*{B}(t)$ is assumed to depend only on time for simplicity. A more detailed explanation of other physical variables is given in the main text.

One can rewrite Eq.\eqref{SEOM1} in a more convenient manner as
\begin{equation}
  \begin{aligned}
    D^\alpha\dot{\vb*{r}}^\alpha&=\tilde{\vb*{v}}^\alpha+\frac{e}{\hbar}\vb*{E}\times\vb*{\Omega}^\alpha+\frac{e}{\hbar}(\tilde{\vb*{v}}^\alpha\vdot\vb*{\Omega}^\alpha)\vb*{B},\\
    D^\alpha\dot{\vb*{p}}^\alpha&=-e\vb*{E}-e\tilde{\vb*{v}}^\alpha\times\vb*{B}-\frac{e^2}{\hbar}(\vb*{E}\vdot\vb*{B})\vb*{\Omega}^\alpha.
  \end{aligned}\label{SEOM2}
\end{equation}

Following the standard approach~\cite{Landau_Kinetics,hydroofelectron_graphene}, the continuity equations for the particle density and the momentum are obtained in the relaxation-time approximation for momentum-relaxing scattering processes as follows:
\begin{align}
  &\partial_tn_{\vb*{B}}+\div\vb*{J}^n_{\vb*{B}}=0,\label{pcl}\\
  &\partial_tP_{\vb*{B},i}+\partial_j\Pi_{\vb*{B},ij}=F_i+\Gamma_i,\label{mcl}
\end{align}
where $n_{\vb*{B}}$ and $\vb*{P}_{\vb*{B}}$ are the particle density and the momentum of electrons under magnetic fields respectively, $\vb*{F}$ is the driving force due to external fields and $\vb*{\Gamma}$ is the momentum-relaxing force. In Eqs.\eqref{pcl} and \eqref{mcl}, $\vb*{J}^n_{\vb*{B}}$ and $\Pi_{\vb*{B},ij}$ are the fluxes of particle density and the momentum.
In the following, we will show the detailed derivation of Eqs.\eqref{pcl} and \eqref{mcl}.

\section{Particle conservation law}
First, multiplying the Boltzmann equation \eqref{bolzmann} with phase-space density of states $D^\alpha(\vb*{p})$ and integrating over the momentum space, we obtain the following equation:
\begin{align}
  \sum_\alpha\int[\dd{\vb*{p}}]D^\alpha\partial_tf^\alpha+\sum_\alpha\int[\dd{\vb*{p}}]D^\alpha\dot{\vb*{r}}^\alpha\vdot\grad_{\vb*{r}}f^\alpha+\sum_\alpha\int[\dd{\vb*{p}}]D^\alpha\dot{\vb*{p}}^\alpha\vdot\grad_{\vb*{p}}f^\alpha=\sum_\alpha\int[\dd{\vb*{p}}]D^\alpha\mathcal{C}[f^\alpha].\label{pcl1}
\end{align}
In the following, we calculate each term respectively. Integrating by parts the first term on the left-hand side, we obtain the expression of the particle density $n_{\vb*{B}}$,
\begin{align}
  \sum_\alpha\int[\dd{\vb*{p}}]D^\alpha\partial_tf^\alpha=\partial_t\sum_\alpha\int[\dd{\vb*{p}}]D^\alpha f^\alpha-\sum_\alpha\int[\dd{\vb*{p}}](\partial_tD^\alpha)f^\alpha=\partial_tn_{\vb*{B}}-\frac{e}{\hbar}\sum_\alpha\int[\dd{\vb*{p}}]\pdv{\vb*{B}}{t}\vdot\vb*{\Omega}^\alpha f^\alpha.\label{pcl2}
\end{align}
Here, $n_{\vb*{B}}\equiv\sum_\alpha\int[\dd{\vb*{p}}]D^\alpha f^\alpha$ is the particle density under magnetic fields. Performing similar procedures for the second and third terms on the left-hand side of Eq.\eqref{pcl1}, we obtain
\begin{align}
  \sum_\alpha\int[\dd{\vb*{p}}]D^\alpha\dot{\vb*{r}}^\alpha\cdot\grad_{\vb*{r}}f^\alpha&=\sum_\alpha\int[\dd{\vb*{p}}]\left(\tilde{v}_i^\alpha+\frac{e}{\hbar}\epsilon_{ijk}E_j\Omega_k^\alpha+\frac{e}{\hbar}(\tilde{v}_j\Omega_j^\alpha)B_i\right)\partial_if^\alpha\nonumber\\
  &=\partial_i\sum_\alpha\int[\dd{\vb*{p}}]\left(\tilde{v}_i^\alpha+\frac{e}{\hbar}\epsilon_{ijk}E_j\Omega_k^\alpha+\frac{e}{\hbar}(\tilde{v}_j\Omega_j^\alpha)B_i\right)f^\alpha-\frac{e}{\hbar}\sum_\alpha\int[\dd{\vb*{p}}]\epsilon_{ijk}(\partial_i E_j)\Omega_k^\alpha f^\alpha\nonumber\\
  &=\div\vb*{J}^N+\frac{e}{\hbar}\sum_\alpha\int[\dd{\vb*{p}}]\pdv{\vb*{B}}{t}\vdot\vb*{\Omega}^\alpha f^\alpha,\label{pcl3}
\end{align}
and
\begin{align}
\sum_\alpha\int[\dd{\vb*{p}}]D^\alpha\dot{\vb*{p}}^\alpha\vdot\grad_{\vb*{p}}f^\alpha&=\sum_\alpha\int[\dd{\vb*{p}}]\left(-eE_i-e\epsilon_{ijk}\tilde{v}^\alpha_jB_k-\frac{e^2}{\hbar}(E_jB_j)\Omega^\alpha_i\right)\partial_{p_i}f^\alpha\nonumber\\
&=\sum_\alpha\int[\dd{\vb*{p}}]\partial_{p_i}\left[\left(-eE_i-e\epsilon_{ijk}\tilde{v}^\alpha_jB_k-\frac{e^2}{\hbar}(E_jB_j)\Omega^\alpha_i\right)f^\alpha\right]\nonumber\\
&\quad-\sum_\alpha\int[\dd{\vb*{p}}]\left(-e\epsilon_{ijk}(\partial_{p_i}\tilde{v}^\alpha_j)B_k-\frac{e^2}{\hbar}(E_jB_j)(\partial_{p_i}\Omega^\alpha_i)\right)f^\alpha\nonumber\\
&=0.\label{pcl4}
\end{align}
Here, we have used the Maxwell equations $\curl\vb*{E}=-\partial_t\vb*{B}$ and $\div\vb*{B}=0$ in the last step of Eq.\eqref{pcl3} and the identity $\epsilon_{ijk}\partial_{p_i}\tilde{v}_j^\alpha=\epsilon_{ijk}\partial_{p_i}\partial_{p_j}\tilde{\epsilon}^\alpha=0$ and $\partial_{p_k}\Omega^\alpha_i=\epsilon_{ijk}\partial_{p_i}\partial_{p_j}A^\alpha_k=0$ in the last step of Eq.\eqref{pcl4}. Here and hereafter, the Einstein summation convention is implied for repeated indices.

For the physical reasons that the total number of particles does not change due to scatterings, both momentum-conserving one and momentum-relaxing one, we conclude that the right-hand side of Eq.\eqref{pcl1} should vanish:
\begin{equation}
  \int[\dd{\vb*{p}}]D^\alpha\mathcal{C}^{\mathrm{mc}}[f^\alpha]=0,\quad\int[\dd{\vb*{p}}]D^\alpha\mathcal{C}^{\mathrm{mr}}[f^\alpha]=0,\quad\mathrm{i.e.}\quad\sum_\alpha\int[\dd{\vb*{p}}]D^\alpha\mathcal{C}[f^\alpha]=0.\label{pcl5}
\end{equation}
Summarizing the above equations~\eqref{pcl1}-\eqref{pcl5}, we obtain the particle conservation law
\begin{equation}
  \partial_tn_{\vb*{B}}+\div\vb*{J}^n_{\vb*{B}}=0,
\end{equation}
owing to the cancelation of $\frac{e}{\hbar}\int[\dd{\vb*{p}}]\pdv{\vb*{B}}{t}\vdot\vb*{\Omega}^\alpha f$ in Eqs.\eqref{pcl2} and \eqref{pcl3}.

\section{Momentum conservation law}
The momentum conserving law is obtained by multiplying the Boltzmann equation~\eqref{bolzmann} with $p_iD^\alpha(\vb*{p})$ and integrating over the momentum space:
\begin{equation}
  \sum_\alpha\int[\dd{\vb*{p}}]p_iD^\alpha\partial_tf^\alpha+\sum_\alpha\int[\dd{\vb*{p}}]p_iD^\alpha\dot{\vb*{r}}^\alpha\vdot\grad_{\vb*{r}}f^\alpha+\sum_\alpha\int[\dd{\vb*{p}}]p_iD^\alpha\dot{\vb*{p}}^\alpha\vdot\grad_{\vb*{p}}f^\alpha=\sum_\alpha\int[\dd{\vb*{p}}]p_iD^\alpha\mathcal{C}[f^\alpha].
\end{equation}
Each term is calculated by taking similar procudures with particle conservation law:
\begin{align}
  \sum_\alpha\int[\dd{\vb*{p}}]p_iD^\alpha\partial_tf^\alpha&=\partial_t\sum_\alpha\int[dp]p_iD^\alpha f^\alpha-\sum_\alpha\int[dp]p_i(\partial_tD^\alpha )f^\alpha\nonumber\\
  &=\partial_tP_{\vb*{B},i}-\frac{e}{\hbar}\sum_\alpha\int[dp]p_i\pdv{\vb*{B}}{t}\vdot\vb*{\Omega}^\alpha f^\alpha,\\
  \sum_\alpha\int[\dd{\vb*{p}}]p_iD^\alpha\dot{\vb*{r}}\vdot\grad_{\vb*{r}}f^\alpha&=\sum_\alpha\int[\dd{\vb*{p}}]p_i\left[\tilde{v}_j^\alpha+\frac{e}{\hbar}\epsilon_{jkl}E_k\Omega_l^\alpha+\frac{e}{\hbar}(\tilde{v}_k^\alpha\Omega_k^\alpha)B_j\right]\partial_jf^\alpha\nonumber\\
  &=\partial_j\sum_\alpha\int[\dd{\vb*{p}}]p_i\left[\tilde{v}_j^\alpha+\frac{e}{\hbar}\epsilon_{jkl}E_k\Omega_l^\alpha+\frac{e}{\hbar}(\tilde{v}_k^\alpha\Omega_k^\alpha)B_j\right]f^\alpha-\frac{e}{\hbar}\epsilon_{jkl}(\partial_jE_k)\sum_\alpha\int[\dd{\vb*{p}}]p_i\Omega_l^\alpha f^\alpha\nonumber\\
  &=\partial_j\Pi_{\vb*{B},ij}+\frac{e}{\hbar}\sum_\alpha\int[\dd{\vb*{p}}]p_i\pdv{\vb*{B}}{t}\vdot\vb*{\Omega}^\alpha f^\alpha,\\
  \sum_\alpha\int[\dd{\vb*{p}}]p_iD^\alpha\dot{\vb*{p}}^\alpha\vdot\grad_{\vb*{p}}f^\alpha&=\sum_\alpha\int[\dd{\vb*{p}}]p_i\left(-eE_j-e\epsilon_{jkl}\tilde{v}^\alpha_kB_l-\frac{e^2}{\hbar}(E_kB_k)\Omega^\alpha_j\right)\partial_{p_j}f^\alpha\nonumber\\
  &=\sum_\alpha\int[\dd{\vb*{p}}]\partial_{p_j}\left[p_i\left\{-eE_j-e\epsilon_{jkl}\tilde{v}^\alpha_kB_l-\frac{e^2}{\hbar}(E_kB_k)\Omega^\alpha_j\right\}f^\alpha\right]\nonumber\\
  &\qquad-\sum_\alpha\int[\dd{\vb*{p}}](\partial_{p_j}p_i)\left[-eE_j-e\epsilon_{jkl}\tilde{v}^\alpha_kB_l-\frac{e^2}{\hbar}(E_kB_k)\Omega^\alpha_j\right]f^\alpha\nonumber\\
  &\qquad-\sum_\alpha\int[\dd{\vb*{p}}]p_i\partial_{p_j}\left[-eE_j-e\epsilon_{jkl}\tilde{v}^\alpha_kB_l-\frac{e^2}{\hbar}(E_kB_k)\Omega^\alpha_j\right]f^\alpha\nonumber\\
  &=\sum_\alpha\int[\dd{\vb*{p}}]\left[eE_i+e\epsilon_{ikl}\tilde{v}_k^\alpha B_l+\frac{e^2}{\hbar}(E_kB_k)\Omega^\alpha_i\right]f^\alpha.
\end{align}
Here, we have defined the total momentum $P_{\vb*{B},i}$ and the momentum flux $\Pi_{\vb*{B},ij}$ as follows:
\begin{align}
  P_{\vb*{B},i}&=\sum_\alpha\int[\dd{\vb*{p}}]p_iD^\alpha f^\alpha,\\
  \Pi_{\vb*{B},ij}&=\sum_\alpha\int[\dd{\vb*{p}}]p_i\left[\tilde{v}_j^\alpha+\frac{e}{\hbar}\epsilon_{jkl}E_k\Omega_l^\alpha+\frac{e}{\hbar}(\tilde{v}_k^\alpha\Omega_k^\alpha)B_j\right]f^\alpha.
\end{align}
After all, they satisfy the momentum conservation law,
\begin{equation}
  \partial_tP_{\vb*{B},i}+\partial_j\Pi_{\vb*{B},ij}=F_i+\Gamma_i
\end{equation}
with the driving force $F_i$ and the momentum-relaxing force $\Gamma_i$ defined as follows respectively:
\begin{align}
  F_i&=\sum_\alpha\int[\dd{\vb*{p}}]\left[eE_i+e\epsilon_{ikl}\tilde{v}_k^\alpha B_l+\frac{e^2}{\hbar}(E_kB_k)\Omega^\alpha_i\right]f^\alpha,\\
\Gamma_i&=\sum_\alpha\int[\dd{\vb*{p}}]p_iD^\alpha\mathcal{C}[f^\alpha]=\sum_\alpha\int[\dd{\vb*{p}}]p_iD^\alpha\mathcal{C}^{\mathrm{mr}}[f^\alpha].
\end{align}
Here, we have used the fact that the integration of the product of the momentum-conserving collision integral $\mathcal{C}^{\mathrm{mc}}[f^\alpha]$ and a fuction of momentum should vanish: $\int[\dd{\vb*{p}}]p_iD^\alpha(\vb*{p})\mathcal{C}^{\mathrm{mc}}[f^\alpha]=0$.

\section{Hydrodynamic variables}
In order to construct the effective hydrodynamic theory, which is correct up to the second order in electric fields and the first order in magnetic fields, we introduce several concepts.
In the hydrodynamic regime, the system reaches the local equilibrium via normal electron-electron scatterings which conserve the total momentum of electron systems. For this reason, we can assume that the distribution functions are described with the Lagrange multiplier $\vb*{u}$ as follows:
\begin{equation}
f_{\mathrm{local}}^\alpha(\vb*{r},\vb*{p},t)=\frac{1}{e^{\beta(\tilde{\epsilon}^\alpha(\vb*{p})-\vb*{u}\vdot\vb*{p}-\mu)}+1},
\end{equation}
which is referred to as the local equilibrium distribution function. Note that the inverse temperature $\beta$, as well as the chemical potential $\mu$ and the drift velocity $\vb*{u}$, generally depend on position and time.

From now on, we assume that the magnetic field is sufficiently weak: $f_{\mathrm{local}}^\alpha=f_{\vb*{u}}^\alpha(\vb*{p})-(\vb*{B}\vdot\vb*{m}^\alpha(\vb*{p}))\partial_\epsilon f_{\vb*{u}}^\alpha(\vb*{p})+O(\vb*{B}^2)$. Here we have introduced the notation
\begin{equation}
  f_{\vb*{u}}^\alpha(\vb*{p})=\frac{1}{e^{\beta(\epsilon_0^\alpha(\vb*{p})-\vb*{u}\vdot\vb*{p}-\mu)}+1}.
\end{equation}
We carry out the calculation under an assumption that the band energy has an isotropic parabolic dispersion with the same effective mass $m$ around some valleys: $\epsilon^\alpha_0=p^2/2m$, where $\vb*{p}$ is defined as a deviation from the valley. For example, when considering graphene with inversion breaking or ML-TMDs, the centers of valleys correspond to $K$ and $K'$ points in the Brillouin zone and the above condition is satisfied.

In the following, we assume that the underlying effective theory is invariant under Galilean transformation:
\begin{equation}
f_{\vb*{u}}^\alpha(\vb*{p}+m\vb*{u})=f_0(\epsilon_0)\label{gt0}
\end{equation}
for a free-like dispersion $\epsilon_0(\vb*{p})=\vb*{p}^2/2m$~\cite{PhysRevResearch.3.013290}. Here, $f_0(\epsilon)\equiv[e^{\beta(\epsilon-\mu_0)}+1]^{-1}$ is the conventional Fermi-Dirac distribution function. Eq.\eqref{gt0} is obtained by transforming the energy and the chemical potential from the frame of reference $\mathcal{S}$ in which the fluid moves with velocity $\vb*{u}$ to the frame $\mathcal{S}_0$ in which it is at rest: $\mu=\mu_0-m\vb*{u}^2/2$.

\subsection{I. particle conservation law}
We are now ready to express two conservation laws in terms of hydrodynamic variables. First, the particle density $n_{\vb*{B}}$ can be expressed as follows:
\begin{align}
  n_{\vb*{B}}&=\sum_\alpha\int[\dd{\vb*{p}}]D^\alpha(\vb*{p})f^\alpha_{\mathrm{local}}\nonumber\\
  &=\sum_\alpha\int[\dd{\vb*{p}}]D^\alpha(\vb*{p})\biggl[f_{\vb*{u}}^\alpha(\vb*{p})-\vb*{B}\vdot\vb*{m}^\alpha(\vb*{p})\partial_\epsilon f_{\vb*{u}}^\alpha(\vb*{p})\biggr]+O(\vb*{B}^2)\nonumber\\
  &=\sum_\alpha\int[\dd{\vb*{p}}]D^\alpha(\vb*{p}+m\vb*{u})\biggl[f_{\vb*{u}}^\alpha(\vb*{p}+m\vb*{u})-\vb*{B}\vdot\vb*{m}^\alpha(\vb*{p}+m\vb*{u})\partial_\epsilon f_{\vb*{u}}^\alpha(\vb*{p}+m\vb*{u})\biggr]+O(\vb*{B}^2)\nonumber\\
  &=\sum_\alpha\int[\dd{\vb*{p}}]\left(1+\frac{e}{\hbar}\vb*{B}\cdot\vb*{\Omega}^\alpha(\vb*{p}+m\vb*{u})\right)f^\alpha_{0}(\epsilon_0)-\vb*{B}\vdot\sum_\alpha\int[\dd{\vb*{p}}]\vb*{m}^\alpha(\vb*{p}+m\vb*{u})\partial_\epsilon f^\alpha_{0}(\epsilon_0)+O(\vb*{B}^2)\nonumber\\
  &=\sum_\alpha\int[\dd{\vb*{p}}]f_0(\epsilon_0)+\frac{e}{\hbar}B_j\sum_\alpha\int[\dd{\vb*{p}}]\frac{\partial\Omega^\alpha_j}{\partial p_i}mu_if_0(\epsilon_0)-B_j\sum_\alpha\int[\dd{\vb*{p}}]\frac{\partial m^\alpha_j}{\partial p_i}mu_i\partial_\epsilon f_0(\epsilon_0)+O(\vb*{B}^2,\vb*{B}\vb*{u}^2)\nonumber\\
  &=n_0+\frac{me}{\hbar}u_iB_j(D_{ij}+N_{ij})+O(\vb*{B}^2,\vb*{B}\vb*{u}^2).\label{pnd}
\end{align}
Here, we have defined the particle number density in the absence of magnetic field
\begin{equation}
  n_0\equiv\sum_\alpha\int[\dd{\vb*{p}}]f_{\vb*{u}}(\vb*{p})=\sum_\alpha\int[\dd{\vb*{p}}]f_{\vb*{u}}(\vb*{p}+m\vb*{u})=\sum_\alpha\int[\dd{\vb*{p}}]f_0(\epsilon_0),
\end{equation}
and one can observe that new geometrical tensors such as Berry curvature dipole $\hat{D}$ ~\cite{Sodemann2015} and $\hat{N}$ appear:
\begin{equation}
  D_{ij}\equiv\sum_\alpha\int[\dd{\vb*{p}}]\frac{\partial\Omega^\alpha_j}{\partial p_i}f_0(\epsilon_0),\quad N_{ij}\equiv-\frac{\hbar}{e}\sum_\alpha\int[\dd{\vb*{p}}]\frac{\partial m_j^\alpha}{\partial p_i}\partial_\epsilon f_0(\epsilon_0).
\end{equation}
Surprisingly, Eq.\eqref{pnd} shows that an interplay between the magnetic field $\vb*{B}$ and the drift velocity $\vb*{u}$ changes the particle number density due to the lack of inversion symmetry.

The same calculations can be performed for the particle number flux $\vb*{J}^n_{\vb*{B}}$ :
\begin{align}
  \vb*{J}^n_{\vb*{B}}=\sum_\alpha\int[\dd{\vb*{p}}]\left[\tilde{\vb*{v}}^\alpha+\frac{e}{\hbar}\vb*{E}\times\vb*{\Omega}^\alpha+\frac{e}{\hbar}(\tilde{\vb*{v}}^\alpha\vdot\vb*{\Omega}^\alpha)\vb*{B}\right]f_{\mathrm{local}}^\alpha.\label{n-flux1}
\end{align}
Now we consider the $i$-component of Eq.\eqref{n-flux1}. Each term can be calculated as follows.
\begin{align}
  \sum_\alpha\int[\dd{\vb*{p}}]\tilde{v}^\alpha_if_{\mathrm{local}}^\alpha&=\sum_\alpha\int[\dd{\vb*{p}}]\tilde{v}^\alpha_i(\vb*{p})\biggl[f_{\vb*{u}}^\alpha(\vb*{p})-\vb*{B}\vdot\vb*{m}^\alpha(\vb*{p})\partial_\epsilon f_{\vb*{u}}^\alpha(\vb*{p})\biggr]+O(\vb*{B}^2)\nonumber\\
  &=\sum_\alpha\int[\dd{\vb*{p}}]\tilde{v}^\alpha_i(\vb*{p}+m\vb*{u})\biggl[f_{\vb*{u}}^\alpha(\vb*{p}+m\vb*{u})-\vb*{B}\vdot\vb*{m}^\alpha(\vb*{p}+m\vb*{u})\partial_\epsilon f_{\vb*{u}}^\alpha(\vb*{p}+m\vb*{u})\biggr]+O(\vb*{B}^2)\nonumber\\
  &=\sum_\alpha\int[\dd{\vb*{p}}](v^\alpha_i(\vb*{p}+m\vb*{u})-B_j\partial_{p_i}m^\alpha_j(\vb*{p}+m\vb*{u}))f_0(\epsilon_0)\nonumber\\
  &\quad-B_j\sum_\alpha\int[\dd{\vb{p}}]v_i^\alpha(\vb*{p}+m\vb*{u})m^\alpha_j(\vb*{p}+m\vb*{u})\partial_\epsilon f_0(\epsilon_0)+O(\vb*{B}^2)\nonumber\\
  &=\sum_\alpha\int[\dd{\vb*{p}}]\left[\frac{p_i+mu_i}{m}-B_j\partial_{p_i}\left(m^\alpha_j(\vb*{p})+\pdv{m^\alpha_j}{p_k}mu_k\right)\right]f_0(\epsilon_0)\nonumber\\
  &\quad-B_j\sum_\alpha\int[\dd{\vb*{p}}]\frac{p_i+mu_i}{m}\left(m_j^\alpha(\vb*{p})+\pdv{m_j^\alpha}{p_k}mu_k\right)\partial_\epsilon f_0(\epsilon_0)+O(\vb*{B}^2,\vb*{B}\vb*{u}^2)\nonumber\\
  &=u_i\sum_\alpha\int[\dd{\vb*{p}}]f_0(\epsilon_0)-B_j\sum_\alpha\int[\dd{\vb*{p}}]\pdv{m_j^\alpha}{p_i}f_0(\epsilon_0)-B_j\sum_\alpha\int[\dd{\vb*{p}}]\frac{p_i}{m}m_j^\alpha(\vb*{p})\partial_\epsilon f_0(\epsilon_0)+O(\vb*{B}^2,\vb*{B}\vb*{u}^2)\nonumber\\
  &=n_0u_i-B_j\sum_\alpha\int[\dd{\vb*{p}}]\pdv{m_j^\alpha}{p_i}f_0(\epsilon_0)-B_j\sum_\alpha\int[\dd{\vb*{p}}]m_j^\alpha\partial_{p_i} f_0(\epsilon_0)+O(\vb*{B}^2,\vb*{B}\vb*{u}^2)\nonumber\\
  &=n_0u_i-B_j\sum_\alpha\int[\dd{\vb*{p}}]\pdv{m_j^\alpha}{p_i}f_0(\epsilon_0)+B_j\sum_\alpha\int[\dd{\vb*{p}}]\pdv{m_j^\alpha}{p_i}f_0(\epsilon_0)+O(\vb*{B}^2,\vb*{B}\vb*{u}^2)\nonumber\\
  &=n_0u_i+O(\vb*{B}^2,\vb*{B}\vb*{u}^2).\\
  \sum_\alpha\int[\dd{\vb*{p}}]\epsilon_{ijk}E_j\Omega^\alpha_kf_{\mathrm{local}}^\alpha&=\epsilon_{ijk}E_j\sum_\alpha\int[\dd{\vb*{p}}]\Omega^\alpha_k(\vb*{p})[f_{\vb*{u}}^\alpha(\vb*{p})-B_lm^\alpha_l(\vb*{p})\partial_\epsilon f_{\vb*{u}}^\alpha(\vb*{p})]+O(\vb*{B}^2)\nonumber\\
  &=\epsilon_{ijk}E_j\sum_\alpha\int[\dd{\vb*{p}}]\Omega^\alpha_k(\vb*{p}+m\vb*{u})[f_{\vb*{u}}^\alpha(\vb*{p}+m\vb*{u})-B_lm_l^\alpha(\vb*{p}+m\vb*{u})\partial_\epsilon f_{\vb*{u}}^\alpha(\vb*{p}+m\vb*{u})]+O(\vb*{B}^2)\nonumber\\
  &=\epsilon_{ijk}E_j\sum_\alpha\int[\dd{\vb*{p}}]\left(\Omega^\alpha_k(\vb*{p})+\pdv{\Omega^\alpha_k}{p_m}mu_m\right)\left[f_0(\epsilon_0)-B_l\left(m_l^\alpha(\vb*{p})+\pdv{m_l^\alpha
  }{p_n}mu_n\right)\partial_\epsilon f_0(\epsilon_0)\right]+O(\vb*{B}^2,\vb*{B}\vb*{u}^2)\nonumber\\
  &=\epsilon_{ijk}E_j\sum_\alpha\int[\dd{\vb*{p}}]\pdv{\Omega^\alpha_k}{p_l}mu_lf_0(\epsilon_0)-\epsilon_{ijk}E_jB_l\sum_\alpha\int[\dd{\vb*{p}}]\Omega^\alpha_km^\alpha_l\partial_\epsilon f_0(\epsilon_0)+O(\vb*{B}^2,\vb*{B}\vb*{u}^2)\nonumber\\
  &=\epsilon_{ijk}mE_j(D_{lk}u_l+Y_{lk}B_l).\\
  \sum_\alpha\int[\dd{\vb*{p}}]\tilde{v}^\alpha_j\Omega^\alpha_jB_if_{\mathrm{local}}^\alpha&=B_i\sum_\alpha\int[\dd{\vb*{p}}]v^\alpha_j(\vb*{p})\Omega^\alpha_j(\vb*{p})f_{\vb*{u}}^\alpha(\vb*{p})+O(\vb*{B}^2)\nonumber\\
  &=B_i\sum_\alpha\int[\dd{\vb*{p}}]v^\alpha_j(\vb*{p}+m\vb*{u})\Omega^\alpha_j(\vb*{p}+m\vb*{u})f_{\vb*{u}}^\alpha(\vb*{p}+m\vb*{u})+O(\vb*{B}^2)\nonumber\\
  &=B_i\sum_\alpha\int[\dd{\vb*{p}}]\frac{p_j+mu_j}{m}\left(\Omega^\alpha_j(\vb*{p})+\pdv{\Omega_j^\alpha}{p_k}mu_k\right)f_0(\epsilon_0)+O(\vb*{B}^2,\vb*{B}\vb*{u}^2)\nonumber\\
  &=B_i\sum_\alpha\int[\dd{\vb*{p}}]\frac{p_j}{m}\Omega^\alpha_jf_0(\epsilon_0)+O(\vb*{B}^2,\vb*{B}\vb*{u}^2)\nonumber\\
  &=B_i\sum_\alpha\int[\dd{\vb*{p}}]\Omega_j^\alpha\partial_{p_j}f_0(\epsilon_0)+O(\vb*{B}^2,\vb*{B}\vb*{u}^2)\nonumber\\
  &=-B_i\sum_\alpha\int[\dd{\vb*{p}}](\partial_{p_j}\Omega_j^\alpha)f_0(\vb*{p})+O(\vb*{B}^2,\vb*{B}\vb*{u}^2)\nonumber\\
  &=O(\vb*{B}^2,\vb*{B}\vb*{u}^2).
\end{align}
In the above calculation, we have used the relation $\partial_{p_i}f_0(\epsilon_0^\alpha)=\pdv{\epsilon_0^\alpha}{p_i}\partial_\epsilon f_0(\epsilon_0^\alpha)=\frac{p_i}{m}\partial_\epsilon f_0(\epsilon_0^\alpha)$ and defined the new geometrical tensor
\begin{equation}
  Y_{lk}=-\frac{1}{m}\sum_\alpha\int[\dd{\vb*{p}}]\Omega_k^\alpha m_l^\alpha \partial_\epsilon f_0(\epsilon_0).
\end{equation}
In the end, we can obtain the particle number flux $\vb*{J}^n_{\vb*{B}}$ as follows,
\begin{align}
  \vb*{J}^n_{\vb*{B}}=n_0\vb*{u}+\frac{me}{\hbar}\vb*{E}\times({}^t\hat{D}\vb*{u}+{}^t\hat{Y}\vb*{B})+O(\vb*{B}^2,\vb*{B}\vb*{u}^2).
\end{align}
Summarizing the above equations, the particle conservation law Eq.~\eqref{pcl} is expressed as
\begin{align}
  \pdv{t}\left[n_0+\frac{me}{\hbar}u_iB_j(D_{ij}+N_{ij})\right]+\div\left[n_0\vb*{u}+\frac{me}{\hbar}\vb*{E}\times({}^t\hat{D}\vb*{u}+{}^t\hat{Y}\vb*{B})\right]=0
\end{align}
up to the order of $\vb*{B}^2$, $\vb*{B}\vb*{u}^2$.

\subsection{II. momentum conservation law}
First we calculate the total momentum:
\begin{align}
  P_{\vb*{B},i}&=\sum_\alpha\int[\dd{\vb*{p}}]p_iD^\alpha(\vb*{p})f^\alpha_{\mathrm{local}}\nonumber\\
  &=\sum_\alpha\int[\dd{\vb*{p}}]p_iD^\alpha(\vb*{p})[f^\alpha_{\vb*{u}}(\vb*{p})-\vb*{B}\vdot\vb*{m}^\alpha(\vb*{p})\partial_\epsilon f_{\vb*{u}}^\alpha(\vb*{p})]+O(\vb*{B}^2)\nonumber\\
  &=\sum_\alpha\int[\dd{\vb*{p}}](p_i+mu_i)D^\alpha(\vb*{p}+m\vb*{u})[f^\alpha_{\vb*{u}}(\vb*{p}+m\vb*{u})-\vb*{B}\vdot\vb*{m}^\alpha(\vb*{p}+m\vb*{u})\partial_\epsilon f_{\vb*{u}}(\vb*{p}+m\vb*{u})]+O(\vb*{B}^2)\nonumber\\
  &=\sum_\alpha\int[\dd{\vb*{p}}](p_i+mu_i)\left(1+\frac{e}{\hbar}\vb*{B}\vdot\vb*{\Omega}^\alpha(\vb*{p}+m\vb*{u})\right)f_{0}(\epsilon_0)-\vb*{B}\vdot\sum_\alpha\int[\dd{\vb*{p}}](p_i+mu_i)\vb*{m}^\alpha(\vb*{p}+m\vb*{u})\partial_\epsilon f_0(\epsilon_0)+O(\vb*{B}^2)\nonumber\\
  &=\sum_\alpha\int[\dd{\vb*{p}}](p_i+mu_i)\left[1+\frac{e}{\hbar}B_j\left(\Omega^\alpha_j(\vb*{p})+\pdv{\Omega^\alpha_j}{p_k}mu_k\right)\right]f_0(\vb*{p})\nonumber\\
  &\quad-B_j\sum_\alpha\int[\dd{\vb*{p}}](p_i+mu_i)\left(m_j^\alpha(\vb*{p})+\pdv{m_j^\alpha}{p_k}mu_k\right)\partial_\epsilon f_0(\epsilon_0)+O(\vb*{B}^2,\vb*{B}\vb*{u}^2)\nonumber\\
  &=mu_i\sum_\alpha\int[\dd{\vb*{p}}]f_0(\epsilon_0)+\frac{e}{\hbar}B_j\sum_\alpha\int[\dd{\vb*{p}}]p_i\Omega_j^\alpha f_0(\epsilon_0)-B_j\sum_\alpha\int[\dd{\vb*{p}}]p_im_j^\alpha\partial_\epsilon f_0(\epsilon_0)+O(\vb*{B}^2,\vb*{B}\vb*{u}^2)\nonumber\\
  &=mu_in_0+\frac{e}{\hbar}B_j(C_{ij}+M_{ij}).
\end{align}
In the third term, we have used the relation for parabolic dispersion $m\partial_{p_i}f_0(\epsilon_0^\alpha)=p_i\partial_\epsilon f_0(\epsilon_0^\alpha)$,
\begin{align}
  -\int[\dd{\vb*{p}}]p_im_j^\alpha\partial_\epsilon f_0(\epsilon_0)=-m\int[\dd{\vb*{p}}]m_j^\alpha\partial_{p_i} f_0(\epsilon_0)=m\int[\dd{\vb*{p}}]\pdv{m_j^\alpha}{p_i}f_0(\epsilon_0).
\end{align}
Here, two new geometrical tensors appear in the expression of the total momentum:
\begin{align}
  C_{ij}=\sum_\alpha\int[\dd{\vb*{p}}]p_i\Omega_j^\alpha f_0(\epsilon_0),\quad M_{ij}=\frac{m\hbar}{e}\sum_\alpha\int[\dd{\vb*{p}}]\frac{\partial m_j^\alpha}{\partial p_i}f_0(\epsilon_0).
\end{align}
Next, we would like to calculate the momentum flux
\begin{align}
  \Pi_{\vb*{B},ij}=\sum_\alpha\int[\dd{\vb*{p}}]p_i\left[\tilde{v}_j^\alpha+\frac{e}{\hbar}\epsilon_{jkl}E_k\Omega_l^\alpha+\frac{e}{\hbar}(\tilde{v}_k^\alpha\Omega_k^\alpha)B_j\right]f^\alpha_{\mathrm{local}}.\label{piij}
\end{align}
The first term of $\Pi_{\vb*{B},ij}$ is calculated as
\begin{align}
\sum_\alpha\int[\dd{\vb*{p}}]p_i\tilde{v}_j^\alpha(\vb*{p}) f^\alpha_{\mathrm{local}}&=\sum_\alpha\int[\dd{\vb*{p}}]p_i\tilde{v}_j^\alpha(\vb*{p})[f_{\vb*{u}}^\alpha(\vb*{p})-\vb*{B}\vdot\vb*{m}^\alpha(\vb*{p})\partial_\epsilon f_{\vb*{u}}^\alpha(\vb*{p})]+O(\vb*{B}^2)\nonumber\\
&=\sum_\alpha\int[\dd{\vb*{p}}](p_i+mu_i)\tilde{v}_j^\alpha(\vb*{p}+m\vb*{u})[f_{\vb*{u}}^\alpha(\vb*{p}+m\vb*{u})-\vb*{B}\vdot\vb*{m}^\alpha(\vb*{p}+m\vb*{u})\partial_\epsilon f_{\vb*{u}}^\alpha(\vb*{p}+m\vb*{u})]+O(\vb*{B}^2)\nonumber\\
&=\sum_\alpha\int[\dd{\vb*{p}}](p_i+mu_i)\left[\frac{p_j+mu_j}{m}-B_k\partial_{p_j}\left(m_k^\alpha(\vb*{p})+\pdv{m_k^\alpha}{p_l}mu_l\right)\right]f_0(\epsilon_0)\nonumber\\
&\quad-B_k\sum_\alpha\int[\dd{\vb*{p}}](p_i+mu_i)\frac{p_j+mu_j}{m}\left(m_k^\alpha(\vb*{p})+\pdv{m_k^\alpha}{p_l}mu_l\right)\partial_\epsilon f_0(\epsilon_0)+O(\vb*{B}^2,\vb*{B}\vb*{u}^2)\nonumber\\
&=\frac{1}{m}\sum_\alpha\int[\dd{\vb*{p}}]p_ip_jf_0(\epsilon_0)+mu_iu_j\sum_\alpha\int[\dd{\vb*{p}}]f_0(\epsilon_0)\nonumber\\
&\quad-B_kmu_i\int[\dd{\vb*{p}}]\pdv{m_k^\alpha}{p_j}f_0(\epsilon_0)-B_k\int[\dd{\vb*{p}}]p_i\frac{\partial^2m_k^\alpha}{\partial p_j\partial p_l}mu_lf_0(\epsilon_0)\nonumber\\
&\quad-B_k\sum_\alpha\int[\dd{\vb*{p}}](p_iu_j+u_ip_j)m_k^\alpha\partial_\epsilon f_0(\epsilon_0)-B_k\sum_\alpha\int[\dd{\vb*{p}}]p_ip_j\pdv{m^\alpha_k}{p_l}u_l\partial_\epsilon f_0(\epsilon_0)
+O(\vb*{B}^2,\vb*{B}\vb*{u}^2).\label{mcl11}
\end{align}
The first term of Eq.~\eqref{mcl11} becomes~\cite{Landau_Kinetics}
\begin{equation}
  \frac{1}{m}\sum_\alpha\int[\dd{\vb*{p}}]p_ip_jf_0(\epsilon_0)=p\delta_{ij}.
\end{equation}
The fourth term of Eq.\eqref{mcl11} can be calculated as
\begin{align}
  -B_k\sum_\alpha\int[\dd{\vb*{p}}]p_i\frac{\partial^2m_k^\alpha}{\partial p_j\partial p_l}mu_lf_0(\epsilon_0)&=-B_k\sum_\alpha\int[\dd{\vb*{p}}]\partial_{p_j}\left(p_i\pdv{m_k^\alpha}{p_l}mu_lf_0(\epsilon_0)\right)+B_k\sum_\alpha\int[\dd{\vb*{p}}](\partial_{p_j}p_i)\pdv{m_k^\alpha}{p_l}mu_lf_0(\epsilon_0)\nonumber\\
  &\quad+B_k\sum_\alpha\int[\dd{\vb*{p}}]p_i\pdv{m_k^\alpha}{p_l}mu_l\partial_{p_j}f_0(\epsilon_0)\nonumber\\
  &=mB_ku_l\delta_{ij}\sum_\alpha\int[\dd{\vb*{p}}]\pdv{m_k^\alpha}{p_l}f_0(\epsilon_0)+B_k\sum_\alpha\int[\dd{\vb*{p}}]p_i\pdv{m_k^\alpha}{p_l}u_lp_j\partial_\epsilon f_0(\epsilon_0)\nonumber\\
  &=\frac{e}{\hbar}B_ku_l\delta_{ij}M_{lk}+B_k\sum_{\alpha}\int[\dd{\vb*{p}}]p_ip_j\pdv{m_k^\alpha}{p_l}u_l\partial_\epsilon f_0(\epsilon).
\end{align}
The sixth term of Eq.\eqref{mcl11} can be calculated as
\begin{align}
  -B_k\sum_\alpha\int[\dd{\vb*{p}}]p_iu_jm_k^\alpha\partial_{\epsilon}f_0(\epsilon_0)+(i\leftrightarrow j)&=-mB_ku_j\sum_\alpha\int[\dd{\vb*{p}}]m_k^\alpha\partial_{p_i}f_0(\epsilon_0)+(i\leftrightarrow j)\nonumber\\
  &=mB_ku_j\sum_\alpha\int[\dd{\vb*{p}}]\pdv{m_k^\alpha}{p_i}f_0(\epsilon_0)+(i\leftrightarrow j)\nonumber\\
  &=\frac{e}{\hbar}B_ku_jM_{ik}+(i\leftrightarrow j).
\end{align}
Finally, we obtain
\begin{align}
  \sum_\alpha\int[\dd{\vb*{p}}]p_i\tilde{v}_j^\alpha f_{\mathrm{local}}^\alpha=mn_0u_iu_j+p\delta_{ij}+\frac{e}{\hbar}B_k(u_lM_{lk}\delta_{ij}+u_jM_{ik})+O(\vb*{B}^2,\vb*{B}\vb*{u}^2).
\end{align}
The rest term of $\Pi_{\vb*{B},ij}$ can be calculated in a similar way.
\begin{align}
  \epsilon_{jkl}E_k\sum_\alpha\int[\dd{\vb*{p}}]p_i\Omega_l^\alpha f_{\mathrm{local}}^\alpha&=\epsilon_{jkl}E_k\sum_\alpha\int[\dd{\vb*{p}}]p_i\Omega_l^\alpha(\vb*{p})[f_{\vb*{u}}^\alpha(\vb*{p})-\vb*{B}\vdot\vb*{m}^\alpha(\vb*{p})\partial_\epsilon f_{\vb*{u}}^\alpha(\vb*{p})]+O(\vb*{B}^2)\nonumber\\
  &=\epsilon_{jkl}E_k\sum_\alpha\int[\dd{\vb*{p}}](p_i+mu_i)\Omega_l^\alpha(\vb*{p}+m\vb*{u})[f_{\vb*{u}}^\alpha(\vb*{p}+m\vb*{u})-\vb*{B}\vdot\vb*{m}^\alpha(\vb*{p}+m\vb*{u})\partial_\epsilon f_{\vb*{u}}^\alpha(\vb*{p}+m\vb*{u})]+O(\vb*{B}^2)\nonumber\\
  &=\epsilon_{jkl}E_k\sum_\alpha\int[\dd{\vb*{p}}](p_i+mu_i)\Omega_l^\alpha(\vb*{p}+m\vb*{u})f_0(\epsilon_0)\nonumber\\
  &\qquad-\epsilon_{jkl}E_kB_n\sum_\alpha\int[\dd{\vb*{p}}](p_i+mu_i)\Omega_l^\alpha(\vb*{p}+m\vb*{u})m_n^\alpha(\vb*{p}+m\vb*{u})\partial_\epsilon f_0(\epsilon_0)+O(\vb*{B}^2)\nonumber\\
  &=\epsilon_{jkl}E_k\sum_\alpha\int[\dd{\vb*{p}}]p_i\Omega_l^\alpha f_0(\epsilon_0)+O(\vb*{B}^2,\vb*{B}\vb*{u}^2,\vb*{u}^3)\nonumber\\
  &=\epsilon_{jkl}E_kC_{il}+O(\vb*{B}^2,\vb*{B}\vb*{u}^2,\vb*{u}^3),\\
  B_j\sum_\alpha\int[\dd{\vb*{p}}]p_i\tilde{v}_k^\alpha\Omega_k^\alpha f_{\mathrm{local}}^\alpha&=B_j\sum_\alpha\int[\dd{\vb*{p}}]p_iv_k^\alpha(\vb*{p})\Omega_k^\alpha(\vb*{p})f_{\vb*{u}}^\alpha(\vb*{p})+O(\vb*{B}^2)\nonumber\\
  &=B_j\sum_\alpha\int[\dd{\vb*{p}}](p_i+mu_i)v_k^\alpha(\vb*{p}+m\vb*{u})\Omega_k^\alpha(\vb*{p}+m\vb*{u})f_{\vb*{u}}^\alpha(\vb*{p}+m\vb*{u})+O(\vb*{B}^2)\nonumber\\
  &=B_j\sum_\alpha\int[\dd{\vb*{p}}](p_i+mu_i)\frac{p_k+mu_k}{m}\Omega_k^\alpha(\vb*{p}+m\vb*{u})f_0(\epsilon_0)+O(\vb*{B}^2)\nonumber\\
  &=B_j\sum_\alpha\int[\dd{\vb*{p}}]p_ip_k\pdv{\Omega^\alpha_k}{p_l}u_lf_0(\epsilon_0)+B_j\sum_\alpha\int[\dd{\vb*{p}}](p_iu_k+u_ip_k)\Omega_k^\alpha f_0(\epsilon_0)+O(\vb*{B}^2,\vb*{B}\vb*{u}^2).\label{46}
\end{align}
To obtain a simpler form, we perform partial integration of the first term on the right-hand side of Eq.~\eqref{46},
\begin{align}
  u_l\sum_\alpha\int[\dd{\vb*{p}}]p_ip_k\pdv{\Omega^\alpha_k}{p_l}f_0(\epsilon_0)&=u_l\sum_\alpha\int[\dd{\vb*{p}}]\partial_{p_l}[\Omega^\alpha_kp_ip_kf_0(\epsilon_0)]-u_l\sum_\alpha\int[\dd{\vb*{p}}]\Omega^\alpha_k\partial_{p_l}[p_ip_kf_0(\epsilon_0)]\nonumber\\
  &=-u_i\sum_\alpha\int[\dd{\vb*{p}}]\Omega^\alpha_kp_kf_0(\epsilon_0)-u_k\sum_\alpha\int[\dd{\vb*{p}}]\Omega^\alpha_kp_if_0(\epsilon_0)-u_l\sum_\alpha\int[\dd{\vb*{p}}]\Omega^\alpha_kp_ip_k\partial_{p_l}f_0(\epsilon_0)\nonumber\\
  &=u_iC_{kk}+u_kC_{ik}-u_l\sum_\alpha\int[\dd{\vb*{p}}]\Omega^\alpha_kp_ip_k\partial_{p_l}f_0(\epsilon_0),
\end{align}
and Eq.~\eqref{46} reduces to
\begin{align}
  B_j\sum_\alpha\int[\dd{\vb*{p}}]p_i\tilde{v}_k^\alpha\Omega_k^\alpha f_{\mathrm{local}}^\alpha&=-B_ju_l\sum_\alpha\int[\dd{\vb*{p}}]\Omega^\alpha_kp_ip_k\partial_{p_l}f_0(\epsilon_0)+O(\vb*{B}^2,\vb*{B}\vb*{u}^2)\nonumber\\
  &=B_ju_lO_{il}+O(\vb*{B}^2,\vb*{B}\vb*{u}^2).
\end{align}
Here we have defined the new geometrical tensor
\begin{equation}
  O_{il}=-\sum_\alpha\int[\dd{\vb*{p}}]p_ip_k\Omega^\alpha_k\partial_{p_l}f_0(\epsilon_0).
\end{equation}
After all, we obtain the following equation for the momenum flux:
\begin{align}
  \Pi_{\vb*{B},ij}&=mn_0u_iu_j+p\delta_{ij}+\frac{e}{\hbar}\epsilon_{jkl}E_kC_{il}+\frac{e}{\hbar}B_k\left(u_lM_{lk}\delta_{ij}+u_jM_{ik}\right)+\frac{e}{\hbar}B_ju_lO_{il}+O(\vb*{B}^2,\vb*{B}\vb*{u}^2).
\end{align}
Driving force is defined as
\begin{align}
  F_i&=-\sum_\alpha\int[\dd{\vb*{p}}]\left[eE_i+e\epsilon_{ikl}\tilde{v}_k^\alpha B_l+\frac{e^2}{\hbar}(E_kB_k)\Omega^\alpha_i\right]f^\alpha_{\mathrm{local}}.
\end{align}
The $i$-component of each term can be calculated as follows:
\begin{align}
  eE_i\sum_\alpha\int[\dd{\vb*{p}}]f_{\mathrm{local}}^\alpha&=eE_i\sum_\alpha\int[\dd{\vb*{p}}][f_{\vb*{u}}^\alpha(\vb*{p})-\vb*{B}\vdot\vb*{m}^\alpha(\vb*{p})\partial_\epsilon f_{\vb*{u}}^\alpha(\vb*{p})]+O(\vb*{B}^2)\nonumber\\
  &=eE_i\sum_\alpha\int[\dd{\vb*{p}}][f_{\vb*{u}}^\alpha(\vb*{p}+m\vb*{u})-\vb*{B}\vdot\vb*{m}^\alpha(\vb*{p}+m\vb*{u})\partial_\epsilon f_{\vb*{u}}^\alpha(\vb*{p}+m\vb*{u})]+O(\vb*{B}^2)\nonumber\\
  &=eE_i\sum_\alpha\int[\dd{\vb*{p}}]f_0(\epsilon_0)-eE_iB_j\sum_\alpha\int[\dd{\vb*{p}}]\left(m_j^\alpha(\vb*{p})+\pdv{m_j^\alpha}{p_k}mu_k\right)\partial_\epsilon f_0(\epsilon_0)+O(\vb*{B}^2,\vb*{B}\vb*{u}^2)\nonumber\\
  &=eE_in_0+O(\vb*{B}^2,\vb*{B}\vb*{u}^2),\\
  e\epsilon_{ikl}B_l\sum_\alpha\int[\dd{\vb*{p}}]\tilde{v}_k^\alpha f_{\mathrm{local}}^\alpha&=e\epsilon_{ikl}B_l\sum_\alpha\int[\dd{\vb*{p}}]\tilde{v}_k^\alpha(\vb*{p})f_{\vb*{u}}^\alpha(\vb*{p})+O(\vb*{B}^2)\nonumber\\
  &=e\epsilon_{ikl}B_l\sum_\alpha\int[\dd{\vb*{p}}]\tilde{v}_k^\alpha(\vb*{p}+m\vb*{u})f_{\vb*{u}}^\alpha(\vb*{p}+m\vb*{u})+O(\vb*{B}^2)\nonumber\\
  &=e\epsilon_{ikl}B_l\sum_\alpha\int[\dd{\vb*{p}}]\frac{p_k+mu_k}{m}f_0(\epsilon_0)+O(\vb*{B}^2,\vb*{B}\vb*{u}^2)\nonumber\\
  &=e\epsilon_{ikl}B_lu_kn_0+O(\vb*{B}^2,\vb*{B}\vb*{u}^2),\\
  E_kB_k\sum_\alpha\int[\dd{\vb*{p}}]\Omega_i^\alpha f_{\mathrm{local}}^\alpha&=E_kB_k\sum_\alpha\int[\dd{\vb*{p}}]\Omega_i^\alpha(\vb*{p})f_{\vb*{u}}^\alpha(\vb*{p})+O(\vb*{B}^2)\nonumber\\
  &=E_kB_k\sum_\alpha\int[\dd{\vb*{p}}]\Omega_i^\alpha(\vb*{p}+m\vb*{u})f_{\vb*{u}}^\alpha(\vb*{p}+m\vb*{u})+O(\vb*{B}^2)\nonumber\\
  &=E_kB_k\sum_\alpha\int[\dd{\vb*{p}}]\left(\Omega_i^\alpha(\vb*{p})+\pdv{\Omega_i^\alpha}{p_j}mu_j\right)f_0(\epsilon_0)+O(\vb*{B^2},\vb*{B}\vb*{u}^2)\nonumber\\
  &=O(\vb*{B^2},\vb*{B}\vb*{u}^2).
\end{align}
After all, the driving force can be calculated as
\begin{equation}
  F_i=-en_0(E_i+\epsilon_{ikl}B_lu_k)+O(\vb*{B}^2,\vb*{B}\vb*{u}^2),\quad\mathrm{i.e.}\quad\vb*{F}=-en_0(\vb*{E}+\vb*{u}\times\vb*{B})+O(\vb*{B}^2,\vb*{B}\vb*{u}^2).
\end{equation}
One recognizes that the driving force $F_i$ corresponds to the conventional Lorentz force for a charged particle.
At this stage, we assume the relaxation time approximation for momentum-relaxating scatterings:
\begin{equation}
  C^{\mathrm{mr}}[f^\alpha]=-\frac{f^\alpha-f_0(\tilde{\epsilon}^\alpha)}{\tau},
\end{equation}
and we can evaluate the relaxing force as
\begin{equation}
  \Gamma_i=\sum_\alpha\int[\dd{\vb*{p}}]p_iD^\alpha(\vb*{p})C^{\mathrm{mr}}[f^\alpha]=-\sum_\alpha\int[\dd{\vb*{p}}]p_iD^\alpha(\vb*{p})\frac{f^\alpha-f_0(\tilde{\epsilon}^\alpha)}{\tau}.\label{55}
\end{equation}
The first term on the right-hand side of Eq.~\eqref{55} can be calculated as
\begin{align}
  \sum_\alpha\int[\dd{\vb*{p}}]p_iD^\alpha(\vb*{p})f_{\mathrm{local}}^\alpha&=\sum_\alpha\int[\dd{\vb*{p}}]p_iD^\alpha(\vb*{p})f_{\vb*{u}}^\alpha(\vb*{p})\nonumber\\
  &=\sum_\alpha\int[\dd{\vb*{p}}](p_i+mu_i)\left(1+\frac{e}{\hbar}\vb*{B}\vdot\vb*{\Omega}^\alpha(\vb*{p}+m\vb*{u})\right)[f_{\vb*{u}}^\alpha(\vb*{p}+m\vb*{u})-\vb*{B}\vdot\vb*{m}^\alpha(\vb*{p}+\vb*{u})\partial_\epsilon f_{\vb*{u}}^\alpha(\vb*{p}+m\vb*{u})]\nonumber\\
  &\quad+O(\vb*{B}^2)\nonumber\\
  &=\sum_\alpha\int[\dd{\vb*{p}}](p_i+mu_i)\left[1+\frac{e}{\hbar}B_j\left(\Omega^\alpha_j(\vb*{p})+\pdv{\Omega_j^\alpha}{p_k}mu_k\right)\right]f_0(\epsilon_0)\nonumber\\
  &\quad-B_j\sum_\alpha\int[\dd{\vb*{p}}](p_i+mu_i)\left[m^\alpha_j(\vb*{p})+\pdv{m^\alpha_j}{p_k}mu_k\right]\partial_\epsilon f_0(\epsilon_0)+O(\vb*{B}^2,\vb*{B}\vb*{u}^2)\nonumber\\
  &=mu_i\sum_\alpha\int[\dd{\vb*{p}}]f_0(\epsilon_0)+\frac{e}{\hbar}B_j\sum_\alpha\int[\dd{\vb*{p}}]p_i\Omega_j^\alpha f_0(\epsilon_0)-B_j\sum_\alpha\int[\dd{\vb*{p}}]p_im^\alpha_j(\vb*{p})\partial_\epsilon f_0(\epsilon_0)+O(\vb*{B}^2,\vb*{B}\vb*{u}^2)\nonumber\\
  &=mu_in_0+\frac{e}{\hbar}B_j\sum_\alpha\int[\dd{\vb*{p}}]p_i\Omega_j^\alpha f_0(\epsilon_0)-B_j\sum_\alpha\int[\dd{\vb*{p}}]p_im^\alpha_j(\vb*{p})\partial_\epsilon f_0(\epsilon_0)+O(\vb*{B}^2,\vb*{B}\vb*{u}^2).
\end{align}
On the other hand, the second term on the right-hand side of Eq.~\eqref{55} can be calculated as
\begin{align}
  \sum_\alpha\int[\dd{\vb*{p}}]p_iD^\alpha(\vb*{p})f_0(\tilde{\epsilon}^\alpha)&=\sum_\alpha\int[\dd{\vb*{p}}]p_iD^\alpha(\vb*{p})[f_0(\epsilon_0)-\vb*{B}\vdot\vb*{m}^\alpha(\vb*{p})\partial_\epsilon f_0(\epsilon_0)]+O(\vb*{B}^2)\nonumber\\
  &=\frac{e}{\hbar}B_j\sum_\alpha\int[\dd{\vb*{p}}]p_i\Omega^\alpha_jf_0(\epsilon_0)-B_j\sum_\alpha\int[\dd{\vb*{p}}]p_im^\alpha_j(\vb*{p})\partial_\epsilon f_0(\epsilon_0)+O(\vb*{B}^2).
\end{align}
Consequently, the momentum-relaxing force $\Gamma_i$ can be expressed in terms of hydrodynamic variables as follows:
\begin{equation}
  \Gamma_i=-\frac{mu_in_0}{\tau}.
\end{equation}

\section{Derivation of Generalized Euler equation}
Combining the particle conservation law Eq.~\eqref{pcl} and the momentum conservation law Eq.~\eqref{mcl}, we can derive the hydrodynamic equation which is a generalization of conventional Euler equation correct up to $\vb*{B}^2$, $\vb*{B}\vb*{u}^2$, $\vb*{u}^3$. First, the particle conservation law
\begin{equation}
  \partial_tn_{\vb*{B}}+\div\vb*{J}^n_{\vb*{B}}=0
\end{equation}
can be expressed by hydrodynamic variables of the order of $O(\vb*{B}^2,\vb*{B}\vb*{u}^2,\vb*{u}^3)$,
\begin{equation}
  \pdv{t}\left[n_0+\frac{me}{\hbar}u_iB_j(D_{ij}+N_{ij})\right]+\partial_i\left[n_0u_i+\frac{me}{\hbar}\epsilon_{ijk}E_j(D_{lk}u_l+Y_{lk}B_l)\right]=0.\label{n0}
\end{equation}
In a similar way, the momentum conservation law
\begin{equation}
  \partial_tP_{\vb*{B},i}+\partial_j\Pi_{\vb*{B},ij}=F_i+\Gamma_i
\end{equation}
can be expressed in terms of hydrodynamic variables as follows:
\begin{align}
  &\pdv{t}\left[mn_0u_i+\frac{e}{\hbar}B_j(C_{ij}+M_{ij})\right]+\partial_j\left[mn_0u_iu_j+p\delta_{ij}+\frac{e}{\hbar}\epsilon_{jkl}E_kC_{il}+\frac{e}{\hbar}B_k(u_lM_{lk}\delta_{ij}+u_jM_{ik})+\frac{e}{\hbar}B_ju_lO_{il}\right]\nonumber\\
  &=-en_0(E_i+\epsilon_{ikl}u_kB_l)-\frac{mn_0u_i}{\tau}.\label{p0}
\end{align}
Substituting Eq.~\eqref{n0} into Eq.~\eqref{p0}, we obtain the generalized Euler equation
\begin{align}
  &mn_0\pdv{u_i}{t}+mn_0(\vb*{u}\vdot\grad)u_i+\partial_ip+en_0(E_i+\epsilon_{ikl}u_kB_l)+\frac{e}{\hbar}M_{ij}\pdv{B_j}{t}+\frac{e}{\hbar}\epsilon_{jkl}E_k\partial_jC_{il}+\frac{e}{\hbar}B_k\partial_j(u_lM_{lk}\delta_{ij}+u_jM_{ik})\nonumber\\
  &+\frac{e}{\hbar}B_j\partial_j(u_lO_{il})=-\frac{mn_0u_i}{\tau}.
\end{align}
If we omit the spatial derivatives of $T$ and $\mu$ for simplicity, we obtain the Euler equation which is shown in the main text:
\begin{align}
  &mn_0\pdv{u_i}{t}+mn_0(\vb*{u}\vdot\grad)u_i+\partial_ip+en_0(E_i+\epsilon_{ikl}u_kB_l)+\frac{e}{\hbar}M_{ij}\pdv{B_j}{t}+\frac{e}{\hbar}B_k(M_{lk}\partial_iu_l+M_{ik}\partial_ju_j)+\frac{e}{\hbar}B_jO_{il}\partial_ju_l\nonumber\\
  &=-\frac{mn_0u_i}{\tau}.
\end{align}

\section{local current}
To relate the hydrodynamic theory with an observable current in optical experiments, we further need to introduce the so-called local current~\cite{Xiao2006} when we consider the magneto-optical responses,
\begin{align}
  \vb*{j}=-e\sum_\alpha\int[\dd{\vb*{p}}]D^\alpha\dot{\vb*{r}}^\alpha f^\alpha+\curl\sum_\alpha\int[\dd{\vb*{p}}]\vb*{m}^\alpha f^\alpha.
\end{align}
The detailed calculation of the second term of the right-hand side is given by
\begin{align}
  \sum_\alpha\int[\dd{\vb*{p}}]m_k^\alpha f^\alpha_{\mathrm{local}}&=\sum_\alpha\int[\dd{\vb*{p}}]m_k^\alpha(\vb*{p}) [f^\alpha_{\vb*{u}}(\vb*{p})-\vb*{B}\vdot\vb*{m}^\alpha(\vb*{p})\partial_\epsilon f_{\vb*{u}}^\alpha(\vb*{p})]+O(\vb*{B}^2)\nonumber\\
  &=\sum_\alpha\int[\dd{\vb*{p}}]m_k^\alpha(\vb*{p}+m\vb*{u}) [f^\alpha_{\vb*{u}}(\vb*{p}+m\vb*{u})-\vb*{B}\vdot\vb*{m}^\alpha(\vb*{p}+m\vb*{u})\partial_\epsilon f_{\vb*{u}}^\alpha(\vb*{p}+m\vb*{u})]+O(\vb*{B}^2)\nonumber\\
  &=\sum_\alpha\int[\dd{\vb*{p}}]\left(m_k^\alpha(\vb*{p})+\pdv{m_k^\alpha}{p_l}mu_l\right)f_0(\epsilon_0)\nonumber\\
  &\quad-B_l\sum_\alpha\int[\dd{\vb*{p}}]\left(m_k^\alpha(\vb*{p})+\pdv{m_k^\alpha}{p_m}mu_m\right)\left(m_l^\alpha(\vb*{p})+\pdv{m_l}{p_n}mu_n\right)\partial_\epsilon f_0(\epsilon_0)+O(\vb*{B}^2,\vb*{B}\vb*{u}^2)\nonumber\\
  &=\sum_\alpha\int[\dd{\vb*{p}}]\pdv{m_k^\alpha}{p_l}mu_lf_0(\epsilon_0)-B_l\int[\dd{\vb*{p}}]m_k^\alpha m_l^\alpha\partial_\epsilon f_0(\epsilon_0)+O(\vb*{B}^2,\vb*{B}\vb*{u}^2).
\end{align}
Therefore, we obtain the local current as follows:
\begin{align}
  \vb*{j}=-en_0\vb*{u}-\frac{me^2}{\hbar}\vb*{E}\times({}^t\hat{D}\vb*{u}+{}^t\hat{Y}\vb*{B})+\frac{e}{\hbar}\grad\times({}^t\hat{M}\vb*{u})-\grad\times(\hat{W}\vb*{B}),
\end{align}
with $W_{kl}=\int[\dd{\vb*{p}}]m_k^\alpha m_l^\alpha\partial_\epsilon f_0(\epsilon_0)$.
In the main text, the local current is described as
\begin{equation}
  \vb*{j}=-en_0\vb*{u}-\frac{me^2}{\hbar}\vb*{E}\times({}^t\hat{D}\vb*{u}+{}^t\hat{Y}\vb*{B})+\frac{e}{\hbar}\grad\times({}^t\hat{M}\vb*{u}),\label{local}
\end{equation}
where the terms with spatial derivatives of $T$ and $\mu$ are omitted for simplicity.

\section{linearization of generalized Euler equation}
Next we move on to the analyses of collective modes. In the study of collective modes, the deviations of the local thermodynamic parameters from their equilibrium values are small. Then, the use of linearized hydrodynamic equations is sufficient. They are obtained by looking for a solution in the form of plane waves, i.e. $\vb*{E}=\tilde{\vb*{E}}e^{i\vb*{k}\vdot\vb*{r}-i\omega t}$ together with similar expressions for other oscillating variables such as $n$, $\vb*{u}$ and $\vb*{B}$. Here and hereafter, the subscript $\vb*{B}$ is omitted for simplicity and the subscript 0 describes the non-oscillating part of physical variables.

First, we linearize the equation of particle conservation law. Substituting $n=n_0+\tilde{n}e^{i\vb*{k}\vdot\vb*{r}-i\omega t}$, $\vb*{u}=\tilde{\vb*{u}}e^{i\vb*{k}\vdot\vb*{r}-i\omega t}$ and $\vb*{B}=\vb*{B}_0+\tilde{\vb*{B}}e^{-i\omega t}$ into the particle conservation law
\begin{equation}
  \pdv{n}{t}+\div\vb*{J}^n=0,
\end{equation}
we obtain the following equation:
\begin{equation}
  \tilde{n}=\frac{\vb*{k}\vdot\tilde{\vb*{u}}}{\omega}n_0+\frac{e}{\hbar}\frac{(\vb*{k}\times\tilde{\vb*{E}})_k}{\omega}Y_{lk}B_0^l.\label{deltaN}
\end{equation}
Next, we linearize the hydrodynamic equation in a similar way and obtain the linearized hydrodynamic equation:
\begin{align}
  &mn_0(-i\omega\tilde{u}_i)+\frac{K}{n_0}i{k}_i\tilde{n}+en_0(\tilde{{E}}_i+\epsilon_{ijk}\tilde{{u}}_j{B}^k_0)-\frac{e}{\hbar}M_{ik}(i\epsilon_{klj}{k}_l\tilde{{E}}_j)+\frac{e}{\hbar}B_0^kM_{lk}\tilde{u}_l(i{k}_i)+\frac{e}{\hbar}(i{k}_j\tilde{{u}}_j)M_{ik}{B}_0^k+\frac{e}{\hbar}(i{k}_k{B}^k_0)O_{ij}\tilde{{u}}_j\nonumber\\
  &=-\frac{mn_0}{\tau}\tilde{{u}}_i,\label{73}
\end{align}
where we use the relation between compressibility $K$ and particle density $n$
\begin{equation}
  K\equiv\frac{1}{\kappa}=-V\left(\pdv{p}{V}\right)=n\pdv{p}{n},\quad\mathrm{i.e.}\quad\grad p=\pdv{p}{n}\grad n=\frac{K}{n}\grad n\simeq\frac{K}{n_0}i\vb*{k}\tilde{n}.
\end{equation}
Substituting $\tilde{n}$ of the form in Eq.\eqref{deltaN} into Eq.\eqref{73}, we obtain
\begin{align}
  &\left[\left(-i\omega+\frac{1}{\tau}\right)\delta_{ij}+\frac{e}{m}\epsilon_{ijk}B_0^k+\frac{e}{\hbar}\frac{iB_0^k}{mn_0}\left(M_{jk}k_i+M_{ik}k_j+O_{ij}k_k\right)+\frac{iK}{mn_0\omega}k_ik_j\right]\tilde{u}_j\nonumber\\
  &=\left[-\frac{e}{m}\delta_{ij}+\frac{e}{\hbar}\frac{iM_{ik}}{mn_0}\epsilon_{jkl}k_l-\frac{e}{\hbar}\frac{iK}{n_0^2\omega}Y_{mk}B_0^m\epsilon_{jkl}k_ik_l\right]\tilde{E}_j.\label{euler}
\end{align}
Solving the above simultaneous equations for $\tilde{\vb*{u}}$ and substituting it into Eq.~\eqref{local}, we obtain the optical conductivity $\hat{\sigma}(\omega,\vb*{k})$.

\section{nonreciprocity in surface magnetoplasmons}
In order to study the nonreciprocal collective modes from hydrodynamic equations, one notices that the inverse compressibility $K\equiv1/\kappa$ is not important for the nonreciprocity because it appears in linearized hydrodynamic equations with $k_ik_j$, so we ignore the term proportional to $K$ in the following analyses. According to symmetry consideration~\cite{Toshio}, the geometrical tensor $\hat{M}$ vanishes in a pseudo-vector constrained in the 2D plane : $M_{ij}=M_i\delta_{jz}$. Here we take the vector $\vb*{M}$ along $y$-axis without loss of generality. The $x$ and $y$ component of Eq.~\eqref{euler} are given by
\begin{equation}
  \left(
  \begin{array}{cc}
    1-i\omega\tau & \omega_{\mathrm{c}}\tau(1+i\zeta q_x)\\
    &\\
    -\omega_{\mathrm{c}}\tau(1-i\zeta q_x) & 1-i\omega\tau+2i\omega_{\mathrm{c}}\tau\zeta q_y
  \end{array}
  \right)
  \left(
  \begin{array}{c}
    \tilde{u}_x\\
    \\
    \tilde{u}_y
  \end{array}\right)=-\frac{e\tau}{m}\left(
  \begin{array}{cc}
    1 & 0\\
    &\\
    i\zeta q_y & 1-i\zeta q_x
  \end{array}
  \right)
  \left(
  \begin{array}{c}
    \tilde{E}_x\\
    \\
    \tilde{E}_y
  \end{array}
  \right),
\end{equation}
respectively with $\zeta=M_y/\hbar n_0$. Solving this equation, we obtain
\begin{align}
  \left(
\begin{array}{c}
  \tilde{u}_x\\
  \\
  \tilde{u}_y
\end{array}\right)&=-\frac{e\tau/m}{Z(\omega,\vb*{q})}\left(
\begin{array}{cc}
  1-i\omega\tau+2i\omega_{\mathrm{c}}\tau\zeta q_y-\omega_{\mathrm{c}}\tau(1+i\zeta q_x)i\zeta q_y & -\omega_{\mathrm{c}}\tau(1+\zeta^2q_x^2)\\
  &\\
  \omega_{\mathrm{c}}\tau(1-i\zeta q_x)+(1-i\omega\tau)i\zeta q_y & (1-i\omega\tau)(1-i\zeta q_x)
\end{array}
\right)
\left(
\begin{array}{c}
  \tilde{E}_x\\
  \\
  \tilde{E}_y
\end{array}
\right).
\end{align}
Here, $Z(\omega,\vb*{q})=(1-i\omega\tau+2i\omega_{\mathrm{c}}\tau\zeta q_y)(1-i\omega\tau)+(\omega_{\mathrm{c}}\tau)^2(1+\zeta^2q_x^2)$. Using the form of local current in terms of hydrodynamic valuables:
\begin{equation}
  \tilde{j}_i=-en_0\tilde{u}_i-\frac{me^2}{\hbar}\epsilon_{ijk}Y_{lk}B_0^l\tilde{E}_j+\frac{e}{\hbar}\epsilon_{ijk}iq_jM_{lk}\tilde{u}_l,
\end{equation}
we obtain the optical conductivity given by
\begin{align}
  \sigma_{xx}&=\sigma_{\mathrm{D}}(1-i\omega\tau)(1+\zeta^2q_y^2)/Z,\\
  \sigma_{xy}&=-\sigma_{\mathrm{D}}(1-i\zeta q_x)[\omega_{\mathrm{c}}\tau(1+i\zeta q_x)+i\zeta q_y(1-i\omega\tau)]/Z-\frac{me^2}{\hbar}Y_{zz}B_0,\\
  \sigma_{yx}&=\sigma_{\mathrm{D}}(1+i\zeta q_x)[\omega_{\mathrm{c}}\tau(1-i\zeta q_x)+i\zeta q_y(1-i\omega\tau)]/Z+\frac{me^2}{\hbar}Y_{zz}B_0,\\
  \sigma_{yy}&=\sigma_{\mathrm{D}}(1-i\omega\tau)(1+\zeta^2q_x^2)/Z.
\end{align}
Here, $\sigma_{\mathrm{D}}$ is the Drude weight and $\vb*{q}=(q_x,q_y)$ is the in-plane wave vector of the incident light.

In the following analysis, we consider surface plasmons in 2D electron hydrodynamic materials such as monolayer graphene. Surface plasmons originate from the collective excitations of electrons coupled to the electromagnetic field at an interface between a dielectric material and a metal. The equation which determines the dispersion relation of surface plasmons is given by
\begin{equation}
f(\omega,\vb*{q})=\left[\frac{\epsilon_1}{\kappa_1}+\frac{\epsilon_2}{\kappa_2}+\frac{i\sigma_{xx}}{\epsilon_0\omega}\right]\left[\kappa_1+\kappa_2-\frac{i\omega\sigma_{yy}}{\epsilon_0c^2}\right]-\frac{\sigma_{xy}\sigma_{yx}}{(\epsilon_0c^2)}=0.\label{smp}
\end{equation}
Here, $\kappa_i=\sqrt{q^2-\epsilon_i(\omega/c)^2}$ $(i=1,2)$, $\epsilon_i$ are the relative dielectric constants in the medium above $(i=1)$ and below $(i=2)$ the graphene, $\epsilon_0$ and $c$ are the dielectric constant and the speed of light in free space respectively.

This equation $f(\omega,\vb*{q})=0$ is difficult to solve analytically with respect to $\omega=\omega(\vb*{q})$, however, we can know the nonreciprocity of the surface magnetoplasmons without solving the equation. Let us consider two different directions of propagation.
\begin{itemize}
  \item If we consider the surface plasmons propagating along the $x$-direction, the in-plane wave vector is $\vb*{q}=(q_x,0)$ and Eq.\eqref{smp} becomes $f(\omega,q_x^2)=0$. This can be solved as $\omega=\omega(q_x^2)$ and we conclude
\begin{equation}
  \omega(q_x)=\omega(-q_x).
\end{equation}

\item If we consider the surface plasmons propagating along the $y$-direction, the in-plane wave vector is $\vb*{q}=(0,q_y)$ and $f(\omega,\vb*{q})$ generally satisfies
\begin{equation}
  f(\omega,q_y)-f(\omega,-q_y)\neq0\label{neq}
\end{equation}
due to the nonreciprocity of $\sigma_{ij}$, i.e., $\sigma_{ij}(\omega,q_y)\neq\sigma_{ij}(\omega,-q_y)$. Substituting $\omega=\omega(-q_y)$ in Eq.\eqref{neq}, we obtain
\begin{equation}
  f(\omega(-q_y),q_y)-f(\omega(-q_y),-q_y)\neq0.\label{86}
\end{equation}
Using the fact that $\omega=\omega(-q_y)$ is the solution of $f(\omega,-q_y)=0$ and satisfies $f(\omega(-q_y),-q_y)=0$, we obtain from Eq.~\eqref{86}
\begin{equation}
  f(\omega(-q_y),q_y)\neq0.
\end{equation}
However, $\omega(q_y)$ satisfies $f(\omega(q_y),q_y)=0$, thus we conclude
\begin{equation}
  \omega(q_y)\neq\omega(-q_y).
\end{equation}
\end{itemize}
In summary, the surface magnetoplasmons acquire nonreciprocity only when they propagate along $\vb*{M}$. On the other hand, the surface magnetoplasmons propagating perpendicular to $\vb*{M}$ are reciprocal.

\bibliographystyle{apsrev4-2}
\bibliography{ref}